\documentclass[12pt]{article}
\pdfoutput=1
\usepackage{float}
\usepackage[final]{graphicx}
\usepackage{amssymb,amsmath}
\usepackage{commath}
\usepackage{epsfig}
\usepackage{epstopdf}
\usepackage{latexsym}
\usepackage{graphicx}
\usepackage{subfigure}
\usepackage{booktabs}
\usepackage{tkz-euclide}
\usepackage{makeidx}
\usepackage{cite}
\usepackage{bm}
\usepackage{geometry}
\usepackage{braket}
\usepackage[margin=20pt,small]{caption}

\usepackage[toc]{appendix}

\usepackage{tikz}
\usetikzlibrary{calc,decorations.markings,arrows.meta,shapes.misc,decorations.pathmorphing,calc,bending}

\usetikzlibrary{arrows.meta,shapes.misc,decorations.pathmorphing,calc,bending}

\tikzset{
  branch point/.style={cross out,draw=black,fill=none,minimum size=2*(#1-\pgflinewidth),inner sep=0pt,outer sep=0pt}, 
  branch point/.default=5
}
\tikzset{
  branch cut/.style={
    decorate,decoration=snake,
    to path={
      (\tikztostart) -- (\tikztotarget) \tikztonodes
    },
    }
  }

\usepackage{color}
\usepackage{datetime}
  
\ifpdf
\fi

\definecolor{cardinal}{rgb}{0.6,0,0}
\definecolor{darkgreen}{rgb}{0,0.5,0}
\definecolor{golden}{rgb}{0.92, 0.7, 0}
\definecolor{midnight}{rgb}{0, 0, 0.5}
\definecolor{darkblue}{rgb}{0.2, 0, 0.8}

\newcommand{\be}{\begin{equation}}
\newcommand{\ee}{\end{equation}}
\newcommand{\bea}{\begin{eqnarray}}
\newcommand{\eea}{\end{eqnarray}}

\begin{document}

\begin{titlepage}

\bigskip
\bigskip
\bigskip
\centerline{{\bf Virasoro algebras, kinematic space  and the spectrum of modular Hamiltonians in CFT$_2$}}
\bigskip
\centerline{ Suchetan Das$^{1,2}$, Bobby Ezhuthachan$^2$, Somnath Porey$^2$, Baishali Roy$^2$}
\bigskip
\bigskip
\bigskip
\centerline{$^1$Department of Physics,}
\centerline{Indian Institute of Technology Kanpur,}
\centerline{Kanpur 208016, India.} 
\bigskip
\bigskip
\centerline{$^2$Ramakrishna Mission Vivekananda Educational and Research Institute,}
\centerline{Belur Math,}
\centerline{Howrah-711202, West Bengal, India.} 
\bigskip
\bigskip
\bigskip
\centerline{suchetan[at]iitk.ac.in,  bobby.ezhuthachan[at]rkmvu.ac.in,}
\centerline{somnathhimu00[at]gm.rkmvu.ac.in, baishali.roy025[at]gm.rkmvu.ac.in  }
%\bigskip
%\centerline{\Large \bf }
%\bigskip
\bigskip
\bigskip
\bigskip

\begin{abstract}

\noindent 
We construct an infinite class of eigenmodes with integer eigenvalues for the Vacuum Modular Hamiltonian of a single interval $N$ in 2d CFT and study some of its interesting properties, which includes its action on OPE blocks as well as its bulk duals. Our analysis suggests that these eigenmodes, like the OPE blocks have a natural description on the so called kinematic space of  CFT$_2$ and in particular realize the Virasoro algebra of the theory on this kinematic space. Taken together, our results hints at the possibility of an effective description of the CFT$_2$ in the kinematic space language.

\end{abstract}

\newpage

%\tableofcontents

\end{titlepage}
\tableofcontents

%\newpage

\rule{\textwidth}{.5pt}\\

%%%%%%%%%%%%%%%%%%%%%%%%%
\section{Introduction}

Research over the past several years has made it abundantly clear that  Quantum information/entropy related ideas play a crucial role  in developing a deeper understanding of Quantum Field Theory and  Quantum Gravity.  The algebraic formulation of QFT (AQFT) in terms of algebra of observables associated to causal domains of spatial subregions \cite{Haag:1992hx},\cite{Witten:2018zxz}, seems to be particularly well suited for such entropic studies. The many successes of this approach include formulating a precise version of various entropy bounds in QFT \cite{Casini:2008cr}-\cite{Bousso:2014uxa}, developing a deeper understanding of RG flows in terms of relative entropy of states,  \cite{Casini:2004bw}-\cite{Casini:2018nym}, proofs of various null energy conditions in QFT \cite{Faulkner:2016mzt}-\cite{Ceyhan:2018zfg}, developing a more precise understanding of bulk Reconstruction \cite{Jafferis:2015del}-\cite{Kabat:2018smf} among others. 

A key role in most of these studies is played by the (total) modular hamiltonian \cite{Casini:2011kv}\footnote{Total modular hamiltonian is defined as the difference of modular hamiltonians of a given subregion and it's complement. In the rest of the note, we refer modular hamiltonian as the total modular hamiltonian.}. In the AQFT formulation, the modular hamiltonian operator $K^{\psi}_{\Sigma}$, for a particular state $|\psi\rangle$ generates an automorphism of the algebra of the operators localized in the causal diamond $\mathcal{D}(\Sigma)$ associated with the spatial subregion $\Sigma$. Under this automorphism, operators localized within $\mathcal{D}(\Sigma)$ transform into each other, thus generating a flow called the modular flow \footnote{ Under this flow, an operator $\mathcal{O}\rightarrow \mathcal{O}(s)$, where $\mathcal{O}(s)\equiv e^{isK}\mathcal{O}e^{-isK}$. Both $\mathcal{O}\;  \textrm{and}\; \mathcal{O}_s\; \textrm{have support within}\; \mathcal{D}_{\Sigma} $}. In applications to holography, the importance of the modular hamiltonian operator comes from its identification, at leading order in the inverse bulk newton's constant($\frac{1}{G_N}$), with the corresponding bulk modular hamiltonian operator, where the corresponding bulk region is the bulk causal diamond associated with the region bounded by the RT surface and $\Sigma$ \cite{Jafferis:2015del}. These modular flows  play an important role in the entanglement wedge reconstruction program\footnote{ Recently, a different, but related, notion of the connes co-cycle flows also have been discussed in the context of extracting bulk information from the entanglement wedge region which is casually disconnected from the boundary \cite{Bousso:2020yxi},\cite{Levine:2020upy}}. It has also been argued that the emergence of a semiclassical bulk spacetime might itself be understood from  the algebra of modular hamiltonians of all subregions in the boundary QFT \cite{Kabat:2018smf}.

Given its relevance, particularly in the context of bulk reconstruction program alluded to above, it would be a useful endeavor to study the modular hamiltonian operator in detail both in general QFTs as well more specifically in simple but concrete examples. One way to characterize these operators would be through the spectrum of its eigenstates. It is reasonable to expect that this spectrum would encode the entanglement content of the QFT. The modular hamiltonian $K^{\psi}_{\Sigma}$ for a state $|\psi\rangle$ and a spatial region  $\Sigma$ annihilates the state, ie $K^{\psi}_{\Sigma}|\psi\rangle =0$. One may then construct its eigenstates by acting on $|\psi\rangle$ with a special class of operators ($\mathcal{O}_{\omega}$) which has the following commutation relation with the modular hamiltonian $[K, \mathcal{O}_{\omega}]=\omega \mathcal{O}_{\omega}$. These are referred to as the modular eigenmodes. It's easy to see that the fourier transform of the modular evolved operators (ie: $\int ds e^{is\omega}\mathcal{O}_{\omega}$) are modular eigenmodes. These eigenmodes, in particular the zero modes play a crucial role in reconstruction of bulk fields inside the entanglement wedge \cite{Kabat:2017mun},\cite{Das:2018ojl}. The zero modes, which commute with the modular hamiltonian  may be thought of as local symmetries of the corresponding state, in the sense that correlation function of operators inside the region $\mathcal{D}(\Sigma)$, would be invariant under transformations (of the operators)  generated by the zero modes. These are local because for the same given state, but for a different region, the modular hamiltonian and therefore the zero modes would be different. It has been argued that in the bulk  these local symmetries generated by the zero modes corresponding to large diffeomorphisms which are not trivial on the RT surface \cite{Czech:2019vih}. Thus the modular eigenmodes seem to have a very important bearing on the emergence of bulk geometry itself.  

Given the above motivations, a detailed study of the modular eigenmodes in these theories would be interesting. While for generic states and regions the modular hamiltonian as well its eigenmodes are nonlocal operators, there are a few examples, where they take a simple form as an integral of local fields. The simplest example of which is the case of the single interval in the vacuum state of a $\textrm{CFT}_2$. In \cite{Das:2019iit}, two of us had shown that OPE blocks of primary fields of different spins are nonzero modular eigenmodes of the modular hamiltonian of the single interval, in the vacuum of CFT$_2$, where the endpoints of the interval corresponds to the location of the two primary fields whose expansion define the OPE block. This generalizes known results in the literature that scalar OPE blocks are modular zero eigenmodes \cite{Czech:2017zfq},\cite{Kabat:2017mun}. 

In this note, we continue with the study initiated in \cite{Das:2019iit}, of the eigenmodes of the vacuum modular hamiltonian for a single interval (labelled as $N$) in 2D CFT. We find a new infinite class of modular eigenmodes with integer eigenvalues and discuss some of their interesting properties.
%Specifically, as we show in section [], these eigenmodes together with the modular hamiltonian satisfy a virasoro algebra. The details of this calculation are presented in appendix B. For obvious reasons we refer to as the modular virasoro algebra (MVA). Moreover as we show in the same section, the new eigenmodes have a very interesting action on the OPE blocks, which are themselves eigenmodes of the modular hamiltonian is quite similar to the action of the usual (local) virasoro generators on the modes of a primary field in CFT$_2$. This fact suggests that the OPE blocks behave as the modes of a primary field under conformal transformations generated by the MVA. Since the OPE blocks can be described naturally as fields living on the so-called kinematic space (k-space), this also suggest that the new eigenmodes have a natural  action on the k-space. We explore the kinematic aspects of this question in subsection[]. Giveniscussion on the bulk dual interpretation of these eigemodes. 
%This draft is organized as follows.  In the next section, after presenting a brief summary of the known examples of modular eigenmodes- the OPE blocks, and their bulk duals, we present the new class of eigenmodes and discuss its interesting properties. 
 The key point we want to make here is that like the OPE blocks, these new eigenmodes we construct here have a natural description on the so-called kinematic space(k-space)\cite{Czech:2016xec}, which is essentially the space of causal diamonds in CFT$_2$. In particular, they realize the virasoro algebra of the CFT$_2$ on this k-space. As evidence of this fact we show that OPE blocks, which are local fields in the k-space description, transform as modes of a primary field under this `k-space virasoro algebra', which we refer to as the modular virasoro algebra(MVA) in the bulk of the text. Moreover, as we show, a subset of the new modular eigenmodes, which generate the global subalgebra of the  MVA representation can be directly identified with the modular hamiltonians of subregions of $N$. We believe that these observations, taken together, hint at the possibility of an equivalent effective description of the CFT$_2$ in the  k-space language, a detailed study of which we leave for future work. 
 
 This draft is organized as follows.  In the next section, after presenting a brief summary of the known examples of modular eigenmodes- the OPE blocks, and their bulk duals, we present the new class of eigenmodes and discuss its interesting properties.  Specifically, we show that these eigenmodes together with the modular hamiltonian satisfy the virasoro algebra.  The details of this calculation are presented in appendix \ref{B}.  We  also compute the commutator of these new class of modular eigenmodes with the  OPE blocks, and show that the result is same as that of the usual (local) virasoro generators with the modes of a primary field in CFT$_2$.  This fact suggests that the OPE blocks transform as modes of a primary field under conformal transformations generated by the MVA. Since the OPE blocks can be described naturally as fields living on the so-called kinematic space(k-space), this also suggest that the new eigenmodes have a natural action on the k-space. We explore the kinematic aspects of this question in subsection \ref{subsec}. 
 
 In section \ref{sec 3}, we focus on the global subalgebra of the MVA. Interestingly, we show that it is isomorphic to the algebra of the modular hamiltonians of $N$ as well as its subregions $N'$ and $N''$ and that they implement the so-called modular inclusion within the lightcone of $N$.  For completeness we review  the  definition  of  the  modular  inclusion  and,  as  an  example,  discuss  the  modular inclusion  for  finite  dimensional  Hilbert  space  in  appendix A.   
 
 In  section  \ref{sec bulk},  we  discuss the bulk dual of our construction. In particular we show the emergence of the RT geodesic very naturally from our constructions. We conclude with a summary of our results as well a discussion on some of the questions and directions opened up in the light of these results, in section \ref{discussion}.

\textbf{Note Added:} While we were in the final stages of preparing this article, \cite{Besken:2020snx} was posted on the arxiv, in which the authors construct Virasoro algebra from generators constructed out of light-ray operators. Although the expression for those operators are similar to what we refer here as `modular Virasoro generators', the context and motivation of the approaches seem to us to be different.
\numberwithin{equation}{section}
\section{Modular Hamiltonian in 2D CFT and its spectrum}
The Modular Hamiltonian of a single interval with endpoints ($z_2$, $z_3$) on a constant time slice ($t=0$) in the vacuum of 2D CFT is given by the following integral expression: 
\begin{equation}
K = \int^{\infty}_{-\infty}d\zeta \frac{(z_{2}-\zeta)(\zeta-z_{3})}{z_{2}-z_{3}}T_{\zeta\zeta}(\zeta) + \int^{\infty}_{-\infty}d\bar{\zeta} \frac{(z_{2}-\bar{\zeta})(\bar{\zeta}-z_{3})}{z_{2}-z_{3}}T_{\bar{\zeta}\bar{\zeta}}(\bar{\zeta})
\end{equation}
Modular eigenmodes are operators which satisfy the following commutation relation with the modular hamiltonian $[ K,\mathcal{O}] = \lambda \mathcal{O}$. In \cite{Kabat:2017mun},\cite{Das:2019iit}, it was shown that global OPE blocks in CFT$_{2}$, are eigen-modes of vacuum modular Hamiltonian. We therefore begin this section, with a brief review of the OPE blocks in $\textrm{CFT}_2$.
%This is one of the few known cases where explicit expressions for the Modular hamiltonian exist. See [] for a incomplete list of others. The Modular Hamiltonian is an important element in the holographic dictionary, primarily due to its identification with its bulk counterpart (ie: $K_{bulk}=K_{bndy}$)[], which then lead to a better understanding of the bulk Reconstruction problem. For these reasons, it is useful and important to study the Modular Hamiltonian in more detail and in particular understand it's spectrum ie :the eigenmodes of the Modular Hamiltonian. An interesting class of  such eigenmodes are provided by the OPE blocks, which are bi-local operators  and, as is clear from the name, is defined via the OPE of two primary fields  as follows.
\subsection{OPE Blocks}
%In \cite{Kabat:2017mun},\cite{Das:2019iit}, it was shown that global OPE blocks in CFT$_{2}$, are eigen-modes of vacuum modular Hamiltonian.
In CFT, a global OPE block $B^{ij}_{k}$ is defined as the contribution of a conformal family (ie a given primary field $\mathcal{O}_k$ of dimension $h_{k}$, $\bar{h}_k$ and all it's global descendants) to the OPE of two  primary operators ($\mathcal{O}_i$, $\mathcal{O}_j$) of dimensions ($h_{i}$, $\bar{h}_{i}$) and ($h_j$, $\bar{h}_{j}$) respectively \cite{Czech:2016xec}. Mathematically, 

\begin{align}\label{opeblock}
\mathcal{O}_{i}(z_{1}, \bar{z}_1)\mathcal{O}_{j}(z_{2}, \bar{z}_2)=z_{12}^{-(h_{i}+h_{j})}\bar{z}_{12}^{-(\bar{h}_{i}+\bar{h}_{j})}\sum_{k}C_{ijk}B_{k}^{ij}(z_{1}, \bar{z}_1; z_{2}, \bar{z}_2)
\end{align}
Here, $C_{ijk}$ is the OPE coefficient, which is dynamical input of the theory. The above equation tells us how the OPE block transforms under global conformal transformations and this is enough to fix the form of the OPE blocks. Indeed, $B^{ij}_{k}$ has an integral expression which can be derived \cite{Czech:2016xec}, \cite{Das:2018ajg} using the shadow operator formalism \cite{Ferrara:1973vz}, \cite{SimmonsDuffin:2012uy} and takes the following form,
\begin{align}\label{ope block}
B^{ij}_{k}(z_{1},\bar{z}_{1};z_{2},\bar{z}_{2}) =& \int_{z_{1}}^{z_{2}}d\zeta \int_{\bar{z}_{1}}^{\bar{z}_{2}}d\bar{\zeta}\left(\frac{(\zeta-z_{1})(z_{2}-\zeta)}{z_{2}-z_{1}}\right)^{h_{k}-1}\left(\frac{z_{2}-\zeta}{\zeta-z_{1}}\right)^{h_{ij}}\times \nonumber \\
&\left(\frac{(\bar{\zeta}-\bar{z}_{1})(\bar{z}_{2}-\bar{\zeta})}{\bar{z}_{2}-\bar{z}_{1}}\right)^{\bar{h}_{k}-1}\left(\frac{\bar{z}_{2}-\bar{\zeta}}{\bar{\zeta}-\bar{z}_{1}}\right)^{\bar{h}_{ij}}\mathcal{O}_{k}(\zeta,\bar{\zeta})
\end{align}
%Given two primary operators $\mathcal{O}_i(x_i)$, $\mathcal{O}_j(x_j)$ in CFT$_2$ the OPE block $\mathcal{B}^k_{ij}(x_i, x_j)$ is a bi-local operator which captures the contribution of the $k^{th}$ global conformal family to the OPE expansion of the two operators.
%\begin{equation}
%\mathcal{O}_i(x)\mathcal{O}_j(y) =\sum_k C_{ijk} \mathcal{B}^k_{ij}()
%\end{equation}
%The above definition fixes the conformal transformation properties of the OPE blocks entirely. In the limit $x_i\rightarrow x_j$, the ope blocks are  In 2D CFTThis and the short distance behaviour of these blocks helps in providing a useful integral representation for these blocks[][], which is given as:
%\begin{equation}
%B^k_{ij} = \int  \mathcal{O}
%\end{equation}

One can now show\footnote{See appendix A of \cite{Das:2019iit}, for the details of the proof.}, using the OPE of $T$, $\bar{T}$ with the primary field $\mathcal{O}$, that these OPE blocks are indeed eigenmodes of $K$, with eigenvalue proportional to the spin difference ($l_{ij}$)of the two operators. 
\begin{equation}
\Big[K, B^{ij}_{k} \Big] = 2\pi il_{ij}B^{ij}_k    
\end{equation}

The scalar zero-modes have been shown to be dual to the so-called geodesic operators \cite{Czech:2016xec}, which are essentially smeared geodesic integrals of the appropriate bulk dual field of $\mathcal{O}_k$. The geodesic endpoints being the location of the two primary fields, whose OPE defines the specific OPE block in question. 
\begin{equation}\label{geodesic op1}
B^{ij}_k = \int_{\lambda} ds\; e^{-s\Delta_{ij}}\phi(x(s), z(s), t(s))    
\end{equation}
Here $\phi$ is the dual scalar field to $\mathcal{O}_k$. ($x,\; t$) are the boundary coordinates while $z$ is the bulk coordinate, and the integral is over a geodesic with end points on the boundary.  The smearing function is $e^{-s\Delta_{ij}}$, with $\Delta_{ij}= \Delta_i -\Delta_j$ being the difference in scaling dimensions of the two operators. This can be generalized to non zero scalar modes \cite{Das:2019iit}, where now the integral is over a Lorentzian cylinder. The cylindrical surface is generated by $K$ and $P_D$ which generate boosts in the plane normal to the geodesic and translations along the geodesic respectively. 
\begin{equation}\label{geodesic op2}
B^{ij}_k = c_k\int_{cylinder} d\tilde{t} ds\; e^{-s(\theta)\Delta_{ij}} e^{-\tilde{t}(\rho,\theta)l_{ij}}\phi(x(\rho,\theta), z(\rho,\theta), t(\rho,\theta)) \end{equation} 
 Here $l_{ij}$ is the spin difference between the two operators, $\tilde{t}$ and $s$ labels the coordinates on the cylinder and the $c_k$ is a normalization constant which can be fixed by the appropriate boundary condition. See \cite{Das:2019iit} for the details of the derivation. In the next section, we introduce the new class of eigenmodes which are smeared integrals of $T$ and $\bar{T}$ and discuss their bulk duals. 
\subsection{A new class of modular eigenmodes and its properties}
%This leads us  naturally to the following generalization for  $L_{n},\bar{L}_{n}$, $\forall n$, such that it reduces to $L_{0,\pm 1},\bar{L}_{0,\pm 1}$ for $n=0,\pm 1$. 
We now present a new class of integrated operators, which are all eigenmodes of the modular hamiltonian. Unlike the OPE blocks, these exist in any 2d CFT and are not theory dependent. As advertised earlier, these also satisfy a virasoro algebra. For this reason, we label them as $\mathbb{L}_n$ and $\bar{\mathbb{L}}_n$. The explicit expressions are given below.
%\left(\frac{z_{2}-z_{1}}{z_{3}-z_{1}}\right)^{n}
%\left(\frac{z_{2}-z_{1}}{z_{3}-z_{1}}\right)^{-n} 
\begin{align}\label{Ln definition}
\mathbb{L}_{n} =a_n\int^{\infty}_{-\infty} d\zeta \;\frac{(z_{2}-\zeta)^{-n+1}(\zeta-z_{3})^{n+1}}{z_{2}-z_{3}} T(\zeta) \\
\bar{\mathbb{L}}_{n} =\bar{a}_n\int^{\infty}_{-\infty} d\bar{\zeta}\; \frac{(z_{2}-\bar{\zeta})^{n+1}(\bar{\zeta}-z_{3})^{-n+1}}{z_{2}-z_{3}}\bar{T}(\bar{\zeta})
\end{align}

Note that naively the integrand in the above formulae, blow up at $z_2$ and $z_3$, however we can regulate the integral, by choosing a deformed contour such that it doesn't pass through $z_2$ and $z_3$. Equivalently, we can give both $z_2$ and $z_3$ a small imaginary component. The $a_n$ are arbitrary normalization constants\footnote {If we choose the normalization constant to be independent of the endpoints $z_2$ and $z_3$, then it is easy to see that the $L_n$ and $\bar{L}_n$ are really only a function of ($z_2$-$z_3)$, however if the normalization constants are non trivial functions of the endpoints, then the eigenmodes are bi-local ($z_2$ and $z_3$)}. In this notation,  $\mathbb{L}_0 +\bar{\mathbb{L}}_0$, with $a_0 =\bar{a}_0 =1$ is the modular hamiltonian of the single interval with endpoints $z_2$ and $z_3$. It can be shown that they satisfies, 
 \begin{equation}
[\mathbb{L}_0, \mathbb{L}_n] = -n\mathbb{L}_n, \; \; [\bar{\mathbb{L}}_0, \bar{\mathbb{L}}_n] = -n \bar{\mathbb{L}}_n      
\end{equation}
It follows that the $\mathbb{L}_n$ and $\bar{\mathbb{L}}_n$ for ($n\neq 0$) are indeed modular eigenmodes. Furthermore, if we normalize the $\mathbb{L}_n$ suitably, which can be done without any loss of generality, such that the $a_n =r^n$ and  $\bar{a}_n =\bar{r}^n$ where $r$ and $\bar{r}$ are two arbitrary constants, then in fact these eigenmodes satisfy the virasoro algebra, with the correct central charge term. 
\begin{align}\label{Virasoro}
[\mathbb{L}_{m},\mathbb{L}_{n}] = (m-n)\mathbb{L}_{m+n} + \frac{c}{12}n(n^{2}-1)\delta_{m+n,0}
\end{align}
For this reason, we refer to this, as the modular virasoro algebra (MVA). As we explain in detail in section \ref{sec 3}, there is a nice geometric interpretation of the global $SO(2,2)$ subalgebra of the MVA. In particular, the generators of this global subalgebra  ie : $\mathbb{L}_{0,\pm}$ and $\bar{\mathbb{L}}_{0,\pm}$ are linear combinations of the holomorphic and antiholomorphic components of the modular hamiltonians corresponding to the subregions $N'(z_1, z_2)$ and $N''(z_3,z_1)$ of $N$. See figure 1.  This is particularly transparent if one parameterizes the normalization constant as follows: $r =\frac{1}{\bar{r}} =\left(\frac{z_{2}-z_{1}}{z_{3}-z_{1}}\right)$. As is clear from figure 1, the $z_1$ in this parametrization is the point within the line segment $N$, which divides it into $N'$ and $N''$. For this reason, in the remainder of the note, we use this normalization for the $\mathbb{L}_n$ and $\bar{\mathbb{L}}_n$.  

Finally we note that there is an interesting 'duality' between the standard generators of the CFT, which we denote as ${\bf L}_n$, and the $\mathbb{L}_n$ we construct here. In particular, there exists a conformal transformation which interchanges the two. The explicit map between the two conformal frames is given in equation \ref{new frame1}. Under this transformation, $\mathbb {L}_n\Longleftrightarrow {\bf L}_n$. In particular,  this means that the modular hamiltonian gets interchanged with the usual CFT hamiltonian.  

%We should point out that the $\mathbb{L}_n$ constructed above are not distinct from the usual local representation of virasoro generators in CFT$_2$.  Indeed, as we show in section (\ref{sec bulk}), there is an explicit conformal transformation,  (equation $\ref{new frame1}$), which maps the two. This transformation specifically maps the modular hamiltonian of any interval $N$ to the generator of dilations in the usual representation. 

As we will argue in the rest of the note, it is natural to interpret the $\mathbb{L}_n$ as realizing the virasoro algebra on the space of causal diamonds in the CFT$_2$, which is termed as the kinematic space (k-space). Evidence of this is provided in the following sections, where we analyze the action of $\mathbb{L}_n$ on the OPE blocks which are bilocal operators in the CFT$_2$ but have a simple local  description in the k-space, and later when we understand the geometric meaning of the global subalgebra of the MVA.
%%%%%%%%%%%%%%%%%%%%%%%%%%%%%%%%%%%%%%%%%%%%%%%%%%%%%%%%%%%%%%%%%%%%%%%%%%%%%%%%%%%%%%%%%%%%%%%%%%%%%%%%%%%%%%%%%%%%%%%%%%%%%%%%%%%%%%%%%%%%%%%%%%%%%%%%%%%%%%%%%%%%%%%%%%%%%%%%%%%%%%%%%%%%%%%%%%%%%%%%%%%%%%%%%%%%%%%%%%%%%%%%%%%%%%%%%%%%%%%%%%%%%%%%%%%%%%%%%%%%%%%%%%%%%%%%%%%%%%%%%%%%%%%%%%%%%%%%%%%%%%%%%%%%%%%%%%%%%%%%%%%%%%%%%%%%%%%%%%%%%%%%%%%%%%%%%%%%%%%%%%%%%%%%%%%%%%%%%%%%%%%%%%%%%%%%%%%%%%%%%%%%%%%%%%%%%%%%%%%%%%%%%%%%%%%%%%%%%%%%%%%%%%%%%%%%%%%%%%%%%%%%%%%%%%%%%%

%%%%%%%%%%%%%%%%%%%%%%%%%%%%%%%%%%%%%%%%%%%%%%%%%%%%%%%%%%%%%%%%%%%%%%%%%%%%%%%%%%%%%%%%%%%%%%%%%%%%%%%%%%%%%%%%%%%%%%%%%%%%%%%%%%%%%%%%%%%%%%%%%%%%%%%%%%%%%%%%%%%%%%%%%%%%%%%%%%%%%%%%%%%%%%%%%%%%%%%%%%%%%%%%%%%%%%%%%%%%%%%%%%%%%%%%%%%%%%%%%%%%%%%%%%%%%%%%%%%%%%%%%%%%%%%%%%%%%%%%%%%%%%%%%%%%%%%%%%%%%%%%%%%%%%%%%%%%%%%%%%%%%%%%%%%%%%%%%%%%%%%%%%%%%%%%%%%%%%%%%%%%%%%%%%%%%%%%%%%%%%%%%%%%%%%%%%%%%%%%%%%%%%%%%%%%%%%%%%%%%%%%%%%%%%%%%%%%%%%%%%%%%%%%%%%%%%%%%%%%%%%%%%%%%%%%%

%A detailed  derivation of these results is presented in appendix \ref{B}. The $z_1$ that appears in the definition of the $L_n$ and $\bar{L}_n$ is a constant, which may be chosen to be any number. Thus this is in fact a one parameter family of eigenmodes. In particular if we take $z_3<z_1<z_2$, as we show later in more detail, the global subalgebra of tha MVA has a direct geometric interpretation as the algebra of modular hamiltonians associated with the subregions of $N$, see figure[].  
\subsection{Action on the OPE blocks}

 We can compute the commutator of the  ``modular" $\mathbb{L}_n$ operators with the OPE blocks, by using the commutator of $T$ with primary operators, which can be obtained from the $T\mathcal{O}$ OPE. 
\begin{align}\label{TO}
[T(\omega),\mathcal{O}_{k}(\zeta,\bar{\zeta})] = 2\pi i(h\partial_{\zeta}\delta(\zeta-\omega)+\delta(\zeta-\omega)\partial_{\zeta})\mathcal{O}_{k}(\zeta,\bar{\zeta}), \\
[\bar{T}(\bar{\omega}),\mathcal{O}_{k}(\zeta,\bar{\zeta})] = -2\pi i(h\partial_{\bar{\zeta}}\delta(\bar{\zeta}-\bar{\omega})+\delta(\bar{\zeta}-\bar{\omega}) \partial_{\bar{\zeta}})\mathcal{O}_{k}(\zeta,\bar{\zeta})
\end{align}
%Using (\ref{TO}) and (\ref{ope block}), the action of $L_{0},\bar{L}_{0}$ is given by,
%\begin{align}
%[L_{0},B^{ij}_{k}] = 2\pi i h_{ij}B^{ij}_{k}\\
%[\bar{L}_{0},B^{ij}_{k}] = -2\pi i \bar{h}_{ij}B^{ij}_{k}
%\end{align}
% Here, we want to compute the action of $L_{n}$, $\bar{L}_{n}$ on global OPE blocks. To do this, first we
One then needs to  evaluate the action of $\mathbb{L}_{n}$ on primary field $\mathcal{O}_{k}(\zeta,\bar{\zeta})$. Using (\ref{Ln definition}), (\ref{TO}) we get
\begin{align}\label{action on O}
[\mathbb{L}_{n},\mathcal{O}_{k}(\zeta,\bar{\zeta})] &= \frac{2\pi i}{z_{2}-z_{3}} (z_{2}-\zeta)^{-n}(\zeta-z_{3})^{n}\left(\frac{z_{21}}{z_{31}}\right)^{n} \nonumber \\
& \times \Big[h_{k}\left(n(z_{2}-z_{3})+(z_{2}+z_{3}-2\zeta)\right) + (z_{2}-\zeta)(\zeta-z_{3})\partial_{\zeta}\Big]\mathcal{O}(\zeta,\bar{\zeta}) \\
[\bar{\mathbb{L}}_{n},\mathcal{O}_{k}(\zeta,\bar{\zeta})] &= \frac{2\pi i}{z_{2}-z_{3}} (z_{2}-\bar{\zeta})^{n}(\bar{\zeta}-z_{3})^{-n}\left(\frac{z_{21}}{z_{31}}\right)^{-n} \nonumber \\
& \times \Big[\bar{h}_{k}\left(-n(z_{2}-z_{3})+(z_{2}+z_{3}-2\bar{\zeta})\right) + (z_{2}-\bar{\zeta})(\bar{\zeta}-z_{3})\partial_{\bar{\zeta}}\Big]\mathcal{O}(\zeta,\bar{\zeta})
\end{align}
%At the present it seems quite complicated to find an useful representation of the algebra constructed out of $L_{n}$s from (\ref{action on O}). However, since $L_{n}$s constitute a bi-local representation of the Virasoro algebra, we may find a simpler or useful action on the bi-local objects in CFT$_{2}$. Such a natural bi-local objects is OPE blocks as we described above. Hence, it is natural to ask about the action of $L_{n}$s on them.  
Using (\ref{ope block}) and (\ref{action on O}), we now present the commutator of the $\mathbb{L}_n$ and the $B^{ij}_k$.
\begin{align}\label{rep}
[\mathbb{L}_{n}, B^{ij}_{k}(z_{2},z_{3})]& = \int_{z_{3}}^{z_{2}}d\zeta \int_{\bar{z}_{3}}^{\bar{z}_{2}}d\bar{\zeta}\left(\frac{(\zeta-z_{3})(z_{2}-\zeta)}{z_{2}-z_{3}}\right)^{h_{k}-1}\left(\frac{z_{2}-\zeta}{\zeta-z_{3}}\right)^{h_{ij}}\times \nonumber \\
&\left(\frac{(\bar{\zeta}-\bar{z}_{3})(\bar{z}_{2}-\bar{\zeta})}{\bar{z}_{2}-\bar{z}_{3}}\right)^{\bar{h}_{k}-1}\left(\frac{\bar{z}_{2}-\bar{\zeta}}{\bar{\zeta}-\bar{z}_{3}}\right)^{\bar{h}_{ij}} \frac{2\pi i}{z_{2}-z_{3}} (z_{2}-\zeta)^{-n}(\zeta-z_{3})^{n}\left(\frac{z_{21}}{z_{31}}\right)^{n}\nonumber \\
& \times \Big[h_{k}\left(n(z_{2}-z_{3})+(z_{2}+z_{3}-2\zeta)\right) + (z_{2}-\zeta)(\zeta-z_{3})\partial_{\zeta}\Big]\mathcal{O}_{k}(\zeta,\bar{\zeta}) \nonumber \\
& = (\text{T.D}) + 2\pi i (nh_{k}-n+h_{ij})\left(\frac{z_{21}}{z_{31}}\right)^{n} \int_{z_{3}}^{z_{2}}d\zeta \int_{\bar{z}_{3}}^{\bar{z}_{2}}d\bar{\zeta}\left(\frac{(\zeta-z_{3})(z_{2}-\zeta)}{z_{2}-z_{3}}\right)^{h_{k}-1} \nonumber \\
&\times \left(\frac{z_{2}-\zeta}{\zeta-z_{3}}\right)^{h_{ij}} \left(\frac{(\bar{\zeta}-\bar{z}_{3})(\bar{z}_{2}-\bar{\zeta})}{\bar{z}_{2}-\bar{z}_{3}}\right)^{\bar{h}_{k}-1}\left(\frac{\bar{z}_{2}-\bar{\zeta}}{\bar{\zeta}-\bar{z}_{3}}\right)^{\bar{h}_{ij}} \left(\frac{z_{2}-\zeta}{\zeta-z_{3}}\right)^{-n}\mathcal{O}_{k}(\zeta,\bar{\zeta}) \nonumber \\
& = 2\pi i \left(\frac{z_{21}}{z_{31}}\right)^{n} [n(h_{k}-1)+h_{ij}]B^{ij-n}_k \; ; \; \text{for} \; \Big(h_{ij}-h_{k} \leq n \leq h_{ij} + h_{k}\Big)
\end{align}
Here (T.D) is the total derivative term which vanishes for $\Big(h_{ij}-h_{k} \leq n \leq h_{ij} + h_{k}\Big)$. 

Equation (\ref{rep}) is identical to the action of usual Virasoro generator
%\footnote{where $l_{n} = \frac{1}{2\pi i} \oint dz z^{n+1}T(z)$.}
$l_{n}$ in CFT$_{2}$ on the modes $\phi_{m}$ of a primary field $\phi$,
\begin{align}\label{primary}
[l_{n},\phi_{m}] = [n(h-1)-m]\phi_{n+m}
\end{align}
with the identification $\phi_{m} =\left(\frac{z_{21}}{z_{31}}\right)^{-h_{ij}}B^{ij}_{k}$. Thus we see that the OPE blocks play the role of modes of some highest weight primary field representation of the MVA. 

To summarize, the key results of this section are: 

\begin{itemize}
    \item[a.] The integrated stress tensor operators $\mathbb{L}_n$ and $\bar{\mathbb{L}}_n$ form an infinite class of modular eigenmodes of the modular hamiltonian corresponding to an single interval $N$ with endpoints ($z_1, z_2$). 
    \item[b.] These modular eigenmodes are bi-local, similar to the OPE blocks, in that they are a function of the end points ($z_1,z_2$) of the interval ($N$).
    \item[c.] Their commutators  satisfy the virasoro algebra. For obvious reasons we refer to this representation of the virasoro algebra as the modular virasoro algebra (MVA). Moreover under a conformal transformation given in equation\ref{new frame1}, the ${\mathbb L}_n\Longleftrightarrow {\bf L}_n$. In particular, the modular hamiltonian is interchanged with the usual CFT hamiltonian. 
    \item[d.] Under the transformations generated by the $\mathbb{L}_n$ and $\bar{\mathbb{L}}_n$, the OPE blocks transforms as should the modes of a primary operator in CFT$_2$.
\end{itemize}

Now, the OPE blocks which are bi-local fields in the CFT$_2$ have a natural description as local fields on the so-called kinematic space (k-space) \cite{Czech:2016xec}-\cite{Czech:2015qta}. In the light of [c] and [d], it is natural to wonder whether there exists an ``effective CFT" description in the k-space itself, with the OPE blocks being the primary fields in this ``k-space cft". We explore the kinematical aspects of this question in the next section.
\subsection{MVA and the Kinematic space}\label{subsec}
%The above results seem to have a nice interpretation on the kinematic space.    
%As first discussed in [], bi-local fields in a CFT such as the OPE blocks, can be visualized as local fields living on the  kinematic space. 
%\subsubsection{A Brief review of the kspace formalism}

The kinematic space of CFT$_2$ is defined as the space of a `pair of space like points' in the CFT. Thus its a four dimensional space, with coordinates given by the coordinates of the two points($t_2, x_2; t_3, x_3$) with signature ($+,-,+,-$). One can fix the metric on this space by demanding its invariance under conformal transformations of both points. This leads to a unique metric.
\begin{equation}\label{kspace metric}
    ds^2_{kspace}= 2\Big[\frac{dz_2dz_3}{(z_2 -z_3)^2} + \frac{d\bar{z}_2d\bar{z}_3}{(\bar{z}_2 -\bar{z}_3)^2}\Big] \textrm{with}\; \;
    z_i = t_i + x_i,\; \bar{z}_i = t_i-x_i 
\end{equation}
Thus the 4d space factorizes into two 2d conformally flat space times, spanned by the two sets of k-space light cone coordinates $(z_2,\; z_3)$ and $(\bar{z}_2,\; \bar{z}_3)$ respectively.   
%On a constant time slice,  spanned by the coordinates ($0,x_1; 0, x_2$), the metric reduces to the following form:
%\begin{equation}
%ds^2_{kspace} = \frac{4}{(x_2 -x_3)^2} dx_2 dx_3
%\end{equation}
%Thus this subspace is a conformally flat space with signature ($+,-$) with $z_2$, $z_3$ being the light cone coordinates.

The  k-space formalism allows us to visualize the OPE blocks, which are bi-local fields in the CFT$_2$ as local fields living on this k-space. Moreover it geometrizes the conformal kinematic properties of these OPE blocks. In particular, this means that the conformal casimir equation which the OPE blocks satisfies, derived from its conformal transformation properties as obtained from its definition in \ref{opeblock}, translates in the k-space terminology into an `equations of motion' to be satisfied by the OPE blocks on the k-space \cite{Czech:2016xec},\cite{deBoer:2016pqk}. For scalar OPE blocks, this is just the Klein-Gordon (KG) equation, while the spinning OPE blocks satisfy a slightly modified KG equation \cite{Das:2018ajg}.
The short distance behaviour of the OPE blocks \footnote{The short distance limit of an ope implies the ope block behaves like a single primary, i.e: \\
 $\lim_{z_{2},\bar{z}_{2}\rightarrow z_{3},\bar{z}_{3}} B_{k}^{ij}(z_{2}, \bar{z}_2; z_{3}, \bar{z}_3) \sim z_{23}^{h_{k}}\bar{z}_{23}^{\bar{h}_{k}}\mathcal{O}_{h_{k},\bar{h}_{k}}(z_{3},\bar{z}_{3})$.} is now interpreted as a boundary condition to be imposed along the k-space coordinates ($ z_{23},\;\bar{z}_{23} \rightarrow 0$)
%in which only the corresponding  primary field contribution to the OPE block survives, is now interpreted as a boundary condition of these local fields in k-space.
%\begin{equation}end{equation}
%Given that the $L_n$ are also function of $z_$ and $z_3$, its natural to expect that these modular eigenmodes Its clear from [], that the $L_n$ are function of $z_2$, $z_3$ and $(z_2+z_3$

\subsubsection{ Realizing $\mathbb{L}_n$ on k-space}

Points [c] and [d] of the last section seem to hint at the possibility of an effective CFT description in the k-space with the OPE blocks being modes of the highest weight representation of the corresponding MVA, which in this interpretation would act locally on the k-space. Now from the explicit form of the $\mathbb{L}_n$, it is clear that they are functions of ($z_2 -z_3$). Now in our case, we had chosen the interval to be on a constant time slice, so that $z_2 = \bar{z}_2$ and $z_3=\bar{z}_3$. If, on the other hand, had we chosen an arbitrary spacelike  interval, then indeed $\mathbb{L}_n$, $\bar{\mathbb{L}}_n$ are function of ($z_{23} = z_2 -z_3$) and ($\bar{z}_{23}=\bar{z}_2 -\bar{z}_3$) respectively. As is clear from the k-space metric \ref{kspace metric}, the $z_{23}$, and $\bar{z}_{23}$ are spatial coordinates on the two decoupled spaces and not light cone coordinates. Thus the $\mathbb{L}_n$ and $\bar{\mathbb{L}}_n$ and should be thought of as generating independent spatial diffeomorphisms along $z_{23}$ and $\bar{z}_{23}$ directions rather than generating conformal transformations in a 2d space. 

It is still possible that there is a useful effective k-space description in terms of product of two  CFT$_1$'s, with a 1d stress tensor which we denote as  $\mathbb{T}(\zeta)$ and $\mathbb{T}(\bar{\zeta})$ given by 
\begin{equation}\label{kspace T}
\mathbb{T}(\zeta) = \sum_n \frac{\mathbb{L}_n(z_{23})}{(\zeta -z_{23})^{n+2}},\; \;  \textrm{and}\; \; \bar{\mathbb{T}}(\bar{\zeta}) = \sum_n \frac{\mathbb{L}_n(\bar{z}_{23})}{(\bar{\zeta}-\bar{z}_{23})^{n+2}}
\end{equation}

such that the OPE blocks are modes of a field $\Phi(\zeta, \bar\zeta)$ with respect to both of the 1d cfts.  Where the field $\Phi$ could be formally mode-expanded in terms of the OPE blocks as follows:
\begin{equation}
\Phi(\zeta, \bar{\zeta}) = \sum_{h_{ij}, \bar{h}_{ij}} \frac{\mathcal{B}^{ij}_{k}(z_2,z_3;\bar{z}_2,\bar{z}_3)}{(\zeta-z_{23})^{h_k-h_{ij}} (\bar{\zeta}-\bar{z}_{23})^{\bar{h}_k -\bar{h}_{ij}}}
\end{equation}

However to establish whether a consistent description of this type can be constructed, would involve proving that it satisfies consistent crossing equations among other things. We do not attempt to answer this question here. The only point we want to make here, is that the algebra of the OPE blocks with the $\mathbb{L}_n$ is consistent with the existence of such an effective k-space description.

Of course, if such an effective description does exist in k-space, it would only be a reformulation of the original 2d cft in the k-space language, and the stress tensors defined via  equation \ref{kspace T}, would also be related to the CFT stress tensor components. Nevertheless, we know that the k-space description is a useful intermediary between the AdS and CFT descriptions, because the k-space has the advantage of being directly identified as the space of bulk geodesics which end on the boundary \cite{Czech:2015qta}. This fact has been used to derive a very simple proof of the identification \ref{geodesic op1},\ref{geodesic op2} of OPE blocks as geodesic operators in AdS \cite{Czech:2016xec}. For these reasons, a way to incorporate cft dynamics in k-space language would be interesting from the AdS/CFT perspective.  We hope to come back to this issue in the near future.

%From While these diffeomorphisms for  generic $n$ looks complicated, for $n=0,\pm 1$, an interesting feature
%\subsubsection{Bulk dual of the modular eigenmodes}
\numberwithin{equation}{section}
\section{The global subalgebra of the MVA}\label{sec 3}

In this section, we focus on the global subalgebra of the MVA, which is spanned by $\mathbb{L}_{0,\pm 1}$ and $\mathbb{\bar{L}}_{0,\pm 1}$. We point out that this subset of  $\mathbb{L}_n$ has a nice geometric interpretation as modular hamiltonians of $N$ itself as well as its subparts, labelled as $N'$ and $N''$. This in turn realizes the action of 'modular inclusion' \cite{Wiesbrock:1992mg}-\cite{Wiesbrock:1996rp}, within $\mathcal{D}_{N}$. For brevity, we refer to this subalgebra as the g-MVA in the rest of the note. We begin with a short discussion on the symmetries of causal diamonds in CFT$_2$.
%\subsection{Algebra of Modular Hamiltonians as a sub algebra of the MLRO algebra}
%\subsection{MLRO and Modular Inclusion}
%\section{Discussion}
%\subsection{g-MVA as the algebra of modular hamiltonians of the subregions of N}\label{global}
\subsection{Symmetries of the CFT$_{2}$ causal diamonds}
%Global conformal symmetries are spacetime isometries of causal diamonds corresponding to spherical region in CFT$_{d}$. 
%In CFT$_{2}$, there is a much richer structure due to the existance of local conformal symmetries. 
CFT$_{2}$ causal diamonds associated with intervals on a constant time slice are preserved by a $SO(1,1)\times SO(1,1)$ subset of global conformal symmetry group $SO(2,2)$. Due to chiral structure of symmetry algebras, one can find a right moving and a left moving conformal killing vector (CKV) which stabilizes the diamond. 
For a causal diamond with upper and lower tips at ($v,\bar{v}$) and ($u,\bar{u}$) respectively (in the light cone coordinates), the  CKVs take the following form,
\begin{align}\label{killingvectors}
K^{\zeta}\partial_{\zeta} = \frac{(v-\zeta)(\zeta-u)}{v-u}\partial_{\zeta} \;, \; K^{\bar{\zeta}}\partial_{\bar{\zeta}} = \frac{(\bar{v}-\bar{\zeta})(\bar{\zeta}-\bar{u})}{\bar{v}-\bar{u}}\partial_{\bar{\zeta}}
\end{align}
Here $\zeta(=X+T),\bar{\zeta}(=X-T)$ are the lightcone coordinates. One can similarly define the corresponding conserved charges as:
\begin{align}\label{killingcharges}
K^{R}=\int^{\infty}_{-\infty}d\zeta \frac{(v-\zeta)(\zeta-u)}{u-v}T_{\zeta\zeta}(\zeta) \;, \; K^{L} = \int^{\infty}_{-\infty}d\bar{\zeta} \frac{(\bar{v}-\bar{\zeta})(\bar{\zeta}-\bar{u})}{\bar{v}-\bar{u}}T_{\bar{\zeta}\bar{\zeta}}(\bar{\zeta})
\end{align}
If we take an interval on $T=0$ slice with endpoints $(z_{2},z_{3})$, the upper and lower tips of the corresponding causal diamond(say $N$) are located at $y^{\mu}=(\frac{z_{2}-z_{3}}{2},\frac{z_{2}+z_{3}}{2})$ and $x^{\mu}=(\frac{z_{3}-z_{2}}{2},\frac{z_{2}+z_{3}}{2})$ respectively. In this case: $(u,\bar{u})\equiv(x^{1}-x^{0},x^{1}+x^{0}) =(z_{2},z_{3})$ and $(v,\bar{v})\equiv (y^{1}-y^{0},y^{1}+y^{0}) =(z_{3},z_{2})$. The total modular Hamiltonian $K_{N}$ for the interval $N$  is the sum of $K^{R}_{N}$ and $K^{L}_{N}$, i.e 
\begin{align}
K = K^{R}_{N} + K^{L}_{N} =\int^{\infty}_{-\infty}d\zeta \frac{(z_{2}-\zeta)(\zeta-z_{3})}{z_{2}-z_{3}}T_{\zeta\zeta}(\zeta) + \int^{\infty}_{-\infty}d\bar{\zeta} \frac{(z_{2}-\bar{\zeta})(\bar{\zeta}-z_{3})}{z_{2}-z_{3}}T_{\bar{\zeta}\bar{\zeta}}(\bar{\zeta})
\end{align}  
This can be derived from the expression of modular Hamiltonian of Rindler half space by a conformal transformation from Rindler wedge to CFT$_{2}$ causal diamond. One can similarly define $P_{D}$ as the antisymmetric combination of $K^R$ and $K^L$, i.e. $P_{D} = K^{R} - K^{L}$ \cite{Czech:2017zfq}. Together, $K$ and $P_{D}$ generates a geometrical flow which preserve the diamond. $K$ generate flows from lower tip to the upper tip, while $P_{D}$ generates flows from left to the right tip of the diamond.

%\subsection{Conformal algebra from the modular Hamiltonians of a set of  three causal diamonds}
%The right and left moving CKVs which are of the form (\ref{killingvectors}), can be expressed as linear combinations of global conformal generators $l_{0}=-\zeta\partial_{\zeta}$, $l_{1}=\zeta^{2}\partial_{\zeta}$, $l_{-1}=\partial_{\zeta}$ and the corresponding left moving generators $\bar{l}_{0}$, $\bar{l}_{1}$ and $\bar{l}_{-1}$. From this it follows that given a set of three causal diamonds, one can establish an isomorphism between the algebra of the corresponding set of ($K$'s and $P_D$'s) and the global conformal algebra generated by the $l_{n}, \bar{l}_{n}(n=0,1,-1)$, as will be shown explicitly below for a specific choice of three causal diamonds.

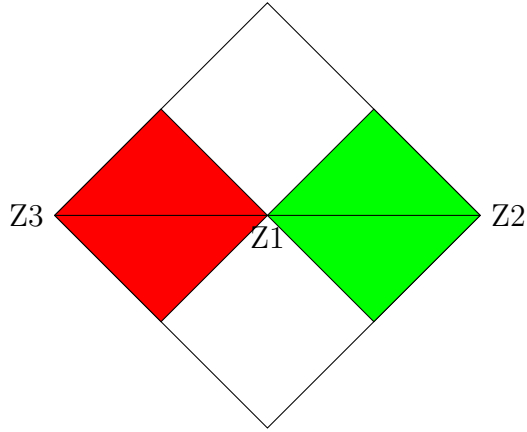
\begin{figure}
\centering
\begin{tikzpicture}
%\draw (0,0) -- (1.41421356237,0);
%\draw (0,0) -- (0.70710678118,0.70710678118);
%\draw (1.41421356237,0) -- (0.70710678118,0.70710678118);
%\draw (1.41421356237,0) -- (2.82842712475,0);
%\draw (1.41421356237,0) -- (2.12132034356,0.70710678118);
%\draw (2.82842712475,0) -- (2.12132034356,0.70710678118);
%\draw (2.82842712475,0) -- (1.41421356237,1.41421356237);
%------------------------------------------------------

\draw  (0,0)  node[anchor= east] {Z3}-- (2.82842712475,2.82842712475) --(5.65685424949,0)  node[anchor=west] {Z2}--(2.82842712475,-2.82842712475) --cycle;
\draw [fill=red](0,0) -- (1.41421356237,1.41421356237)--(2.82842712475,0) node[anchor=north] {Z1}--(1.41421356237,-1.41421356237)--cycle;
\draw [fill=green](2.82842712475,0)--(4.24264068712,1.41421356237)--(5.65685424949,0)--(4.24264068712,-1.41421356237)--cycle;
\draw (0,0)--(5.65685424949,0);

%------------------------------------------------------
%\draw (2.41421356237,0) -- 
%\draw (4,0) -- (135:2cm);
\end{tikzpicture}
\caption{Causal diagram of different regions on a $T=0$ slice.} \label{fig1}
\end{figure}

Consider a CFT$_{2}$ interval $N(z_{3},z_{2})$ on a time slice $T=0$\footnote{i.e. $z_{3,2} = \bar{z}_{3,2}$}. Divide this line segment into two parts $N'(z_{1},z_{2})$ and $N''(z_{3},z_{1})$ around a point $z_{1}$. The corresponding causal diamonds of $N'$ and $N''$ divides the causal diamond of $N$ into four parts such that ($N\supset{N',N'',U,L}$), where $U$ and $L$ are the upper and lower diamond as shown in the Figure(\ref{fig1})\footnote{We will be using the same labels interchangeably for the line segments as well as the corresponding causal diamonds.}.
%\vskip 1 cm
%\vskip 1 cm
%Using the spacetime representation of CKVs as in (\ref{killingvectors}) or using the corresponding charges as in (\ref{killingcharges}) and $TT$ OPE \footnote{For the general derivation of algebras from the $TT$ commutator, we should refer the reader to look at appendix (\ref{B})}, One can work out the algebras of the $K^L$ of them for the diamonds $N,N',N'',U,L$. In particular, we will see the following two subsets of CKV generators(or charges), consisting of three diamonds $N,N',N''$ and $U,L,N'$, form $SO(2,2)$ algebra separately. 
Using the  $TT$ OPE, it can be shown that the $K_{(N',N,N'')}^{R}$ satisfy the following algebra:
\begin{equation}
[K_{N'}^{R},K_{N}^{R}]=2\pi i\left(K_{N}^{R}-K_{N'}^{R}\right)
\end{equation}
\begin{equation}
[K_{N}^{R},K_{N''}^{R}]=2\pi i\left(K_{N}^{R}-K_{N''}^{R}\right)
\end{equation}
\begin{equation}
[K_{N'}^{R},K_{N''}^{R}]=-2\pi i\left(K_{N'}^{R}+K_{N''}^{R}\right)
\end{equation}

This is isomorphic to the holomorphic $SO(2,1)$ subsector of the full $SO(2,2)$ conformal algebras, with the following identifications:\footnote{We have absorbed the $2\pi i$ factor by redefining $K^{R,L}$ as $\frac{1}{2\pi i}K^{R,L}$.}
\begin{equation}\label{right algebra}
K_{N''}^{R}-K_{N}^{R}=\mathbb{L}_{1}\; ; \; K_{N}^{R}=\mathbb{L}_{0}\; ; \; K_{N'}^{R}-K_{N}^{R}=\mathbb{L}_{-1}
\end{equation}

Similarly, $K_{(N',N,N'')}^{L}$ can be shown to satisfy the following commutation relations:
\begin{equation}
[K_{N'}^{L},K_{N}^{L}]=2\pi i\left(K_{N'}^{L}-K_{N}^{L}\right)
\end{equation}
\begin{equation}
[K_{N}^{L},K_{N''}^{L}]=2\pi i\left(K_{N''}^{L}-K_{N}^{L}\right)
\end{equation}
\begin{equation}
[K_{N'}^{L},K_{N''}^{L}]=2\pi i\left(K_{N'}^{L}+K_{N''}^{L}\right)
\end{equation}
This is again isomorphic to the anti-holomorphic $SO(2,1)$ sub-algebra, with the identification:
\begin{equation}\label{left algebra}
K_{N'}^{L}-K_{N}^{L}=\bar{\mathbb{L}}_{1}\; ; \; K_{N}^{L}=\bar{\mathbb{L}}_{0} \; ; \; K_{N''}^{L}-K_{N}^{L}=\bar{\mathbb{L}}_{-1}
\end{equation}
%To distinguish this representation of the conformal algebra from the $l_n$'s, we refer to the $L_n$'s as the 'modular conformal generators'.
The remaining right and left chiral generators of the diamond $U$ and $L$ can be expressed in terms of $K^{R,L}_{N,N',N''}$ as follows:
\begin{align}
K^{R}_{U}=K^{R}_{N''} \; , \; K^{L}_{U}=K^{L}_{N'} \; , \; K^{R}_{L}=K^{R}_{N'} \; , \; K^{L}_{L}=K^{L}_{N''}
\end{align}
Thus, the six modular generators of the three diamonds, ($K^{R(L)}_{N,U,L}$s) also satisfy the $SO(2,1)\times SO(2,1)$ global conformal algebra. 

We denote the CKV's associated with these generators $\mathbb{L}_{0,\pm 1}$ as $L_{0,\pm}$. It's easy to see that these are simply linear combinations of the standard representations of the global conformal generators $l_{1,0,-1}$ defined earlier. The exact relation between them are given by:
%Collectively, it provides a new basis of conformal algebra in terms of . We denote such algebra as `modular conformal algebra'.
%\vskip 0.2 cm
%\textbf{Modular conformal generators as a basis:}
%\vskip 0.2 cm
%In the standard basis around $\zeta,\bar{\zeta}=0$,  as mentioned previously.
%It is easy to see that the modular conformal generators $\mathbb{L}_{1,0,-1}$ and $\bar{\mathbb{L}}_{1,0,-1}$ can be recast as a linear combinations of the 
%$l_n$'s in the following way:
\begin{align}\label{relation between basis}
l_{1} = \frac{2z_{2}z_{3}}{z_{3}-z_{2}}L_{0} + \frac{z_{3}^{2}(z_{1}-z_{2})}{(z_{1}-z_{3})(z_{3}-z_{2})}L_{-1} + \frac{z_{2}^{2}(z_{1}-z_{3})}{(z_{1}-z_{2})(z_{3}-z_{2})}L_{1} \\
l_{0} = \frac{(z_{2}+z_{3})}{z_{2}-z_{3}}L_{0} + \frac{z_{3}(z_{1}-z_{2})}{(z_{1}-z_{3})(z_{2}-z_{3})}L_{-1} + \frac{z_{2}(z_{1}-z_{3})}{(z_{1}-z_{2})(z_{2}-z_{3})}L_{1} \\
l_{-1} = \frac{2}{z_{3}-z_{2}}L_{0} + \frac{(z_{1}-z_{2})}{(z_{1}-z_{3})(z_{3}-z_{2})}L_{-1} + \frac{(z_{1}-z_{3})}{(z_{1}-z_{2})(z_{3}-z_{2})}L_{1}
\end{align}
With a similar set of relations between $\bar{l}_{1,0,-1}$ and $\bar{L}_{1,0,-1}$. 
%Since for any interval in CFT$_{2}$ vacuum, total modular Hamiltonian $K$ and $P_{D}$ can be expressed as a linear combination of conformal generators $l_{0,-1,1}$, we may use (\ref{relation between basis}) to write them in the basis of modular conformal generators. Hence, we could construct the similar algebra as we get, out of the CKVs of any three causal diamonds. However 
As is clear from the definitions, in our set-up, $K_{N}=\mathbb{L}_{0}+\bar{\mathbb{L}}_{0}$ and $\mathbb{L}_{1,-1}$ and $\bar{\mathbb{L}}_{1,-1}$ are it's $\pm 1$ eigenmodes. These eigenmodes are constructed by the $K$s and $P_{D}$s of the regions which reside inside the $N$ itself. This set-up exhibits some other features like modular inclusion as we discuss in the next subsection.
\subsection{g-MVA and modular inclusions}
Out of the left and right moving CKVs, one can notice that $K_{N}$, $K_{U}$, $K_{L}$ and $P_{D,N}$, $P_{D,N'}$, $P_{D,N''}$ are closed under $SO(2,1)$ subalgebra separately\footnote{However these two $SO(2,1)$ don't commute with each other.}.
\begin{align}\label{p inclusion}
[P_{D,N},P_{D,N'}] = 2\pi i(P_{D,N'} - P_{D,N}) \\
[P_{D,N},P_{D,N''}] = 2\pi i(P_{D,N} - P_{D,N''}) \\
[P_{D,N''},P_{D,N'}] = 2\pi i(P_{D,N''}+P_{D,N'}) 
\end{align}
Hence, $P_{D,N''}-P_{D,N}$, $P_{D,N}$ and $P_{D,N'}-P_{D,N}$ satisfy the $SO(2,1)$ sub-algebra.

In a similar fashion, one could also obtain the following
\begin{align}\label{k inclusion}
[K_{N},K_{U}] = K_{N} - K_{U} \\
[K_{N},K_{L}] = K_{L} - K_{N} \\
[K_{U},K_{L}] = K_{U} + K_{L}
\end{align} 
Here $K_{U} - K_{N}$, $K_{N}$ and $K_{L} - K_{N}$ construct the another $SO(2,1)$ sub-algebra. 

The above commutation relations (\ref{p inclusion}) and (\ref{k inclusion}) has the structure of modular inclusion as we discuss in detail in the appendix (\ref{A}). In particular, (\ref{p inclusion}) gives an unitary geometric operation using which one can find a map of algebra of observable between the nested diamonds $N$, $N'$ and $N''$. Similarly, using (\ref{k inclusion}) we have a map of algebra of observables between $N$, $U$ and $L$. See appendix (\ref{A}), for further details. Using the inclusion properties and the fact that $K$ and $P_{D}$ of any diamond can be constructed in the basis of modular generators of $N$, $N'$, $N''$ as in (\ref{relation between basis}), we could in principle construct algebra of observables of any region (diamond) or provide a map to any diamond in the spacetime from $N$. However, since here the modular generators in vacuum are constructed out of conformal symmetries, this inclusion property i.e.: the map from different causal domains is just an artefact of the global conformal symmetry.
\numberwithin{equation}{section}
\section{Pulling the $\mathbb{L}_n$ into the bulk}\label{sec bulk}

Given the explicit form of the $\mathbb{L}_n$ and $\bar{\mathbb{L}}_n$, one can read off the corresponding CKV's ( $L_n$) from equations \ref{Ln definition} and \ref{killingvectors}. Their explicit forms are as follows.  
\begin{eqnarray}
&& L_{n} = \frac{(z_{2}-\zeta)^{-n+1}(\zeta-z_{3})^{n+1}}{z_{3}-z_{2}}\left(\frac{z_{2}-z_{1}}{z_{3}-z_{1}}\right)^{n} \partial_{\zeta},\nonumber \\ 
&& \bar{L}_{n} = -\frac{(z_{2}-\bar{\zeta})^{n+1}(\bar{\zeta}-z_{3})^{-n+1}}{z_{3}-z_{2}}\left(\frac{z_{3}-z_{1}}{z_{2}-z_{1}}\right)^{n} \partial_{\bar{\zeta}}
\end{eqnarray}

Following \cite{Anand:2017dav}, in this section we will extend the CKV's into the bulk where they would generate bulk diffeomorphisms. From our previous discussion of section \ref{sec 3}, we already know what these are for the special case of $n=0,\pm 1$. Since the $\mathbb{L}_{0,\pm 1}$ and $\bar{\mathbb{L}}_{0,\pm 1}$ generate isometries of the causal diamond of the subregions $N'$ and $N''$, their duals would generate boosts around and translations along the respective RT geodesics for $N'$ and $N''$.     

For the generic $n$ case, we proceed as follows. We first use the following transformation $(\zeta,\bar{\zeta}) \rightarrow (\zeta',\bar{\zeta'})$ to transform $(L_{n},\bar{L}_{n}) \rightarrow (l_{n},\bar{l}_{n})$. 
\begin{equation}\label{new frame1}
\zeta' = \frac{1}{\beta}(\frac{\zeta -z_3}{z_2 -\zeta}), \; \; \;\bar{\zeta}' =\beta\frac{z_2 -\bar{\zeta}}{\bar{\zeta}-z_3},\;  (\textrm{with}\; \beta = \frac{z_{13}}{z_{21}})
%\bar{\zeta'} = \frac{z_{31}}{z_{21}} - \frac{1}{\alpha'(\bar{\zeta} - z_{3})} \; ; \; \alpha' = \frac{z_{21}}{z_{23}z_{31}}
\end{equation}

One can then extend these transformations into the bulk. ($y$, $\zeta$, $\bar{\zeta}$) $\rightarrow$ ($y'$, $\zeta'$, $\bar{\zeta'}$). Working in the Fefferman Graham gauge \cite{Banados:1998gg}, the corresponding dual bulk transformations are given by: 
%The bulk duals of the $l_n$'s would be the usual large diffeomorphisms. One can then find the bulk duals of the $L_n$ by using the transformation given in eqn (\ref{\largediff}). 
%Thus in the bulk extension of the $\zeta'$ coordinates, the $L_n$ take the form of the local virasoro generators $l_n$, it Thus to find the bulk representation of the $L_n$, we need to find the 
%If we choose $\alpha,\alpha' = 1$, we could get the $SL(2,\mathbb{R}_\times SL(2,\mathbb{R})$ transformation\footnote{This choice makes the generators $L_{n}$s free from the dependence of $z_{1}$, but it constraints the choice of $z_{2},z_{3}$ also. However, for the purpose to study the bulk dependence we can choose such special point $z_{2,3}$ which makes our life simpler.}. Hence, we can determine the bulk counterpart of this transformation in a straightforward way. Let us choose the most common 
%For any boundary Virasoro transformation $(z,\bar{z}) \rightarrow (g(z),\bar{g}(\bar{z}))$, the bulk diffeomorphism preserving the Fefferman Graham gauge, takes the following form,
%\begin{align}\label{largediff}
%&z'_{b} = g(z) - \frac{2y^{2}(g'(z))^{2}\bar{g}''(\bar{z})}{4g'(z)\bar{g}'(\bar{z})+y^{2}g''(z)\bar{g}''(\bar{z})} \nonumber \\
%&\bar{z}'_{b} = \bar{g}(\bar{z}) - \frac{2y^{2}(\bar{g}'(\bar{z}))^{2}g''(z)}{4g'(z)\bar{g}'(\bar{z})+y^{2}g''(z)\bar{g}''(\bar{z})} \\
%y' = y\frac{4(g'(z)\bar{g}'(\bar{z}))^{\frac{3}{2}}}{4g'(z)\bar{g}'(\bar{z})+y^{2}g''(z)\bar{g}''(\bar{z})}
%\end{align}
%Putting (\ref{new frame1}) in the above transformation we get,
\begin{eqnarray}\label{largediff}
&&\zeta' = \frac{1}{\beta}[\frac{\zeta -z_3}{z_2 -\zeta} - \frac{z_{23}}{z_2 -\zeta} \frac{y^2}{y^2 -(z_2 -\zeta)(\bar{\zeta} -z_3)}] \nonumber \\
&&\bar{\zeta}' = \beta[\frac{z_2 -\bar{\zeta}}{\bar{\zeta} - z_3} - \frac{z_{23}}{\bar{\zeta} -z_3} \frac{y^2}{y^2 -(z_2 -\zeta)(\bar{\zeta} -z_3)}] \nonumber \\
&& y' = \frac{y}{y^2 -(z_2 -\zeta)(\bar{\zeta} -z_3)}
\end{eqnarray}
In the $y\rightarrow 0$ limit, this equation reduces to  equation ($\ref{new frame1}$), as it should.  From the above transformations, we can now obtain the expression for the bulk counterpart of the $L_n$'s. In the primed coordinates, the action of the bulk $l_n$, is given by \cite{Anand:2017dav} 
\begin{equation}\label{primedll}
 l^{(b)}_n = \delta_n\zeta' \partial_{\zeta'} + \delta_n\bar{\zeta'}\partial_{\bar{\zeta'}} +\delta_n y'\partial_{y'}
\end{equation}
where:
\begin{equation}\label{primedl}
\delta_n \zeta' = (-\zeta')^{n+1}, \; \delta_n\bar{\zeta'} = -n(n+1)y'^2(-\zeta')^{n-1} ,\; \delta_n y = \frac{1}{2}(n+1)y'(-\zeta')^{n}
\end{equation}
Similar expression may be obtained for the $\bar{l}^b_n$. By using eqn($\ref{largediff}$) and eqn($\ref{primedl}$) in eqn($\ref{primedll}$), we can obtain the explicit expression for $L^b_n$
\begin{equation}
L^b_n =  \frac{{(-\zeta')}^{n+1}}{z_{21} .A .RT}[U\partial_{\zeta} + V\partial_{\bar{\zeta}} + W\partial_{y}] 
\end{equation}
Here, the expression for $\zeta'$ is given by the first of the equations(\ref{largediff}) and the explicit expressions of $U$, $V$, $W$, $A$ and $RT$ are given below.
\begin{eqnarray}
&&U= y^4 n(n+1) -A(A+z_{21}(n+1)y^2),\nonumber\\ 
&& V=-A(A+z_{21}(n+1)(z_2 -\zeta)(\bar{\zeta}-z_3)) +n(n+1)(\bar{\zeta}-z_3)^2(z_2 -\zeta)^2, \nonumber\\
&& W=  A(2A+z_{21}((\bar{\zeta}-z_3)(z_2-\zeta)+y^2)) +2ny^2(\bar{\zeta}-z_3)(n-z_2+\zeta) \nonumber \\
&& + nz_{21}[(z_2 -\zeta)^2(\bar{\zeta}-z_3)^2(\zeta-z_3)+y^2(y^2(z_2 +z_3-2\zeta)+(z_2 -\zeta)^2(\bar{\zeta}-z_3))],\nonumber \\
&& A= (\bar{\zeta}-z_3)(\zeta-z_3)(z_2-\zeta) +y^2(z_2 +z_3 -2\zeta), \nonumber \\
&& RT=(y^2 - (z_2 -\zeta)(\bar{\zeta} -z_3)) 
\end{eqnarray}

 An interesting feature of the above formulae is the emergence of the RT geodesic. For instance, notice the appearance of the RT geodesic expression ($y^2 -(z_2 -\zeta)(\bar{\zeta} -z_3) $) in the RHS of eqn($\ref{largediff}$). Thus these equations blow up on the RT surface. This means that the bulk coordinates $(\zeta, \bar{\zeta}, y)$  cover only the region between the geodesic and the boundary. Thus they provide a natural set of coordinates for the entanglement wedge associated to $N$. The RT geodesic also appears in the expressions for the bulk counterparts of $L_n$'s given in equation ($\ref{primedll}$). The fact that the bulk coordinates and the bulk extensions of the $L_n$ `know' about the RT geodesic is not surprising. It is simply a reflection of the fact that the boundary $L_n$ are modular eigenmodes by construction and thus has information about the boundary causal diamond of $N$.

\section{Discussion}\label{discussion}
In summary, we have constructed an infinite class of modular eigenmodes ($\mathbb{L}_n$) for the single interval in the vacuum of CFT$_2$. These are expressed as smeared integrals of the stress tensor components and thus exist in any CFT$_2$ \footnote{Such smeared intergrals of the stress tensor has appeared in several different contexts recently, for instance in the study of the light ray operators \cite{Kravchuk:2018htv}-\cite{Besken:2020snx} as well in the context of the so-called dipolar quantization of CFT$_2$ as discussed in \cite{Ishibashi:2015jba},\cite{Ishibashi:2016bey}. We thank Bartek Czeck for bringing these works on the dipolar quantization to our attention.}.
%We further showed that they satisfy  a Virasoro algebra, thus providing a basis for the conformal transformations of the CFT$_2$. 
 Our construction of these eigenmodes are intimately tied to the causal diamond of $N$. This fact manifests itself in many of its interesting features. For instance, one way in which this connection to the causal diamond manifests itself is in the way $\mathbb{L}_n$ acts on OPE blocks. We showed that this action is identical to the action of conformal generators on local primary fields in CFT$_2$.  Coupled with the fact that the OPE blocks have a local description as fields living on the k-space, which is the space of causal diamonds of the CFT$_2$, this  hints at the possibility of finding an equivalent effective description of the CFT on k-space. We argued that on this k-space, the $\mathbb{L}_n$ seem to generate 1d diffeomorphisms along two independent directions. Unfortunately our discussions are only at a kinematic level, and it would be nice if these ideas can be made more concrete. 

The connection to the causal diamonds is even more transparent, in the subclass of the eigenmodes corresponding to $n=0,\pm 1$. In fact, as we showed, these generators are essentially linear combinations of the modular hamiltonians of the causal diamonds for the subregions $N'$ and $N''$ of $N$. We further showed how this structure of the g-MVA realizes modular inclusions within in this setup. The half sided modular inclusion has been studied previously in some examples like certain regions on null plane in higher dimensions \cite{Casini:2017roe} and it has been used to show that in certain special situation, black hole interior could be reconstructed from the algebra of exterior region \cite{Jefferson:2018ksk}. In our example, the inclusion structure emerges quite naturally due to the rich symmetric structure of the vacuum\footnote{As a testing ground of such algebraic structure or modular properties it is always very useful to study them in quantum mechanical system having finite dimensional Hilbert space\cite{Witten:2018zxz}. With this motivation in mind, in appendix \ref{A} we study inclusion properties in an example of finite dimensional Hilbert space where inclusion algebras are satisfied trivially.}. 

Finally we also discussed the action of the bulk counterparts of the $\mathbb{L}_n$ on the bulk spacetime. We saw that these dual descriptions already `know' about the bulk RT geodesic, which is again a reflection of the close connection of our construction with the causal diamond.

A natural question that arises is whether one can extend this construction of algebra and its representation beyond the vacuum in CFT for at least some class of excited states\footnote{For locally excited states in CFT$_{2}$ which are connected to vacuum by local conformal transformation, we do have local expression for modular Hamiltonian in single interval \cite{Das:2018ojl}, \cite{Cardy:2016fqc}. However, we expect this case to be almost identical to the vacuum case.}. Perhaps a more tractable direction to pursue would be to find the extension of such algebras for disconnected multi-interval cases where analytic expression of modular Hamiltonian are known\cite{Casini:2009vk},\cite{Erdmenger:2020nop}.\footnote{Recently analytic expression of modular Hamiltonian for intervals in BMS invariant field theories has been discussed where we could study similar construction to study algebra\cite{Apolo:2020qjm}.} 
%Another plausible direction  would be to show whether one could find BMS algebras from algebras of vacuum modular Hamiltonians for intervals in BMS invariant field theories\cite{Apolo:2020qjm}.  

 We hope to return to some of these questions in the near future.

%In this note, we coonstructed a
%Summary,Discussion of inclusion in higher dimension(Casini paper)),finite case,any doable things beyond the vacuum? excited state? Doing same thing in the excited state which is related to vacuum by local conformal transformation(which includes thermal state also): can we get similar result?Connection to Virasoro OPE block? what is $L_{n}(n \geq 2)$s in terms of modular Hamiltonian? connection to the results with czech paper? connection to modular chaos? $L_{n}$s are modular scrambling modes? 

\vspace*{1ex}
\noindent{\bf Acknowledgment:} 
SD would like to acknowledge the support provided by the Max Planck Partner Group grant MAXPLA/PHY/2018577. The work of SP and BR was supported by a Junior Research Fellowship(JRF) from UGC. 

\appendix
\numberwithin{equation}{section}
\section{Modular inclusion in CFT$_{2}$ and finite dimensional system}\label{A}
\subsection{Modular inclusion}
The Reeh-Schileder theorem states \footnote{The readers may look at \cite{Witten:2018zxz} for a recent review on algebraic QFT and modular theory} that an algebra $\mathcal{A}_{V}$, made out of bounded operators restricted to an arbitrary small open set $V$ in spacetime (flat), is enough to generate (by acting on vacuum) the full vacuum sector of the Hilbert space. Due to this property the vacuum state is said to be `cyclic' w.r.t the algebra of operators $\mathcal{A}_{V}$ in that small open region $V$. Incorporating microcausality, an obvious conclusion can be drawn that such a state is also separating w.r.t $\mathcal{A}_{V}$, which means that, there exists no operator in $V$ which annihilate the vacuum. In algebraic QFT, some useful quantum information quantities like Relative entropy, total modular Hamiltonian can be rigorously constructed for such cyclic and separating states of the QFT Hilbert space. In particular, a self-adjoint `modular operator' $\Delta$ ($=e^{-K}$, K is the total modular Hamiltonian) and an antiunitary operator `modular conjugation' $J$ are the central objects of `Tomita-Takesaki theory', which lies at the foundation of modular theory or modular algebra. The main result of the Tomita-Takesaki theory is that $\Delta$ defines an automorphism which maps an algebra of a region to itself while $J$ defines an isomorphism from the algebra to its commutant $\mathcal{A}'_{V}$.
\begin{align}
\Delta^{is}A\Delta^{-is} = \tilde{A} \; ; \; JAJ = A' ,\; \; \; (A,\tilde{A}) \in \mathcal{A}_{V}, \; A' \in \mathcal{A}'_{V}, \; \forall s \in \mathbb{R}
\end{align}
The $\Delta$ generates a modular flow w.r.t total modular Hamiltonian $K$. Here, the algebra $\mathcal{A}_{V}$ is considered to be a type of Von-Neumann algebra such that, $\mathcal{A}_{V} = \hat{\mathcal{A}}_{V}$. Where, $\hat{\mathcal{A}}_{V}$ is the algebra of the causal domain of the region $V$.

Within the context of the Tomita-Takesaki theory, a notion of inclusion of algebras has been discussed -the so-called `half-sided modular inclusion' (hsmi)\cite{Borchers:1991xk}-\cite{Borchers:2000pv}. Take two Von-Neumann algebra of observables $M,M'$, such that the vacuum $\Omega$ is a common cyclic and separating state for both of them. We can define $\tilde{M} \subset M$ as the +hsmi if it satisfies the condition that $M'$ is preserved under the modular flow of $M$, i.e
\begin{align}\label{+hsmi}
\Delta^{-it}_{M}\tilde{M}\Delta^{it}_{M} \subset \tilde{M}, \; \; \; \forall t \geq 0
\end{align}
Here $\Delta_{M},\Delta_{\tilde{M}}$ are modular operator of $M,\tilde{M}$. \footnote{The corresponding modular conjugation operators are $J_{M},J_{\tilde{M}}$. However, in the present context, we won't need the properties of $J$s and we only focus on modular flows generated by $\Delta$s. For further  details, we refer the readers to the following references \cite{Wiesbrock:1992mg},\cite{Borchers:2000pv}.}
Once the above condition is satisfied, one can construct an one-parameter unitary group $U(a)$ on the Hilbert space such that,
\begin{align}
U(a) = e^{iap}; \; p \equiv \frac{1}{2\pi}(\ln\Delta_{\tilde{M}}-\ln\Delta_{M}) \geq 0; \; \forall a \in \mathbb{R}
\end{align}
The generator $p$ is a positive operator. In such settings, the following properties hold:
\begin{align}
\Delta_{M}^{it}U(a)\Delta_{M}^{-it} &= \Delta_{\tilde{M}}^{it}U(a)
\Delta_{\tilde{M}}^{it} = U(e^{-2\pi t}a); \; \forall a,t \in \mathbb{R} \\
\Delta_{\tilde{M}}^{it} &= U(1) \Delta^{it}_{M}U(-1); \; \forall t \in \mathbb{R} \\
\tilde{M} &= U(1)MU(-1)\\
\Delta_{M}^{it}\Delta_{\tilde{M}}^{-it} &= e^{i\left(-1+e^{-2\pi t}\right)p}
\end{align}
One can see that the first two relations are solved by 
\begin{align}\label{inclusion equation}
[K_{M},K_{\tilde{M}}] = 2\pi i p; \; K_{M,\tilde{M}} = -\ln\Delta_{M,\tilde{M}}
\end{align}
We get the last two relations from the first two. Hence, if $\tilde{M} \subset M$ is a modular inclusion, then (\ref{inclusion equation}) must be satisfied. When the condition of inclusion (\ref{+hsmi}) is satisfied for $t \leq 0$, it is called -hsmi. For that case, the commutation relation of modular Hamiltonian is given by $[K_{M},K_{\tilde{M}}] = -2\pi i p$. Using this $\pm$hsmi, a representation of the  $SL(2,\mathbb{R})$  could be constructed in the following way \cite{Wiesbrock:1996rq},\cite{Wiesbrock:1996rp}
\vskip 0.2 cm
\textbf{Theorem:}\\
\textit{Let $M,M_{1},M_{2}$ be Von-Neumann algebras on a Hilbert space $\mathcal{H}$ and $\Omega$ is a cyclic and separating state $\Omega \in \mathcal{H}$. Assume:
\begin{itemize}
\item{1} $M_{1} \subset M$ is a -hsmi
\item{2} $M_{2} \subset M$ is a +hsmi
\item{3} $M_{2} \subset M'_{1}$ is a -hsmi 
\end{itemize}
(where $M'_{1}$ is the commutant of $M_{1}$.) Then 
$\Delta_{M}^{it}, \Delta_{M_{1}}^{ir}, \Delta^{is}_{M_{2}}$ , $t,r,s \in \mathbb{R}$ 
generate a representation of  $SL(2,\mathbb{R})$ where,
\begin{align}
P \equiv \frac{1}{2\pi}\left(\ln\Delta_{M_{1}}-\ln\Delta_{M}\right); \; K \equiv \frac{1}{2\pi}\left(\ln\Delta_{M_{2}}-\ln\Delta_{M}\right); \; D \equiv \frac{1}{2\pi}\ln\Delta_{M}
\end{align}
}In this way, the algebraic structure of modular inclusion provides an interesting way to construct chiral part of 2D conformal algebra.
\subsection{Modular inclusion in vacuum CFT$_{2}$}
Within the set up of section \ref{sec 3}, we can explicitly see (\ref{p inclusion}) and (\ref{k inclusion}) exhibits both $\pm$hsmi structure. However, (\ref{p inclusion}) is constructed out of $P_{D}$ which is not the modular Hamiltonian. 
%Since, $P_{D}$s are constructed out of modular generators that keep the corresponding diamonds invariant, we could identify them as an identical candidate of modular Hamiltonian $K$. 
However, we can see $P_{D}$s of $N$, $N'$, $N''$ satisfies all the criterion of hsmi. Hence, in CFT$_{2}$ vacuum, we define two types of inclusion structure which we call `$K$-inclusion' and `$P_{D}$-inclusion'. 
\vskip 0.2 cm
\textbf{$P_{D}$-inclusion}\\
Let us first consider the two nested diamonds $N'$ and $N$ where $N' \subset N$. Since $P_{D}$ generates a geometric flow from the left tip to the right tip of a diamond, the algebra of smaller nested diamond $N'$ remain invariant under the flow of $P_{D}$ of the larger diamond $N$ i.e. $e^{-iP_{D,N}t}N'e^{iP_{D,N}t} \subset N'$. In such case, we call such inclusion $N' \subset N$ as the `$P_{D}$-inclusion' which satisfies (\ref{+hsmi}). From the algebra of (\ref{p inclusion}) we have seen that $P_{D,N'}$ and $P_{D,N}$ indeed satisfy half sided modular inclusion algebra  which is $[P_{D,N},P_{D,N'}] = 2\pi i(P_{D,N'} - P_{D,N})$.
Since $P_{D}$ is self adjoint, we could construct a self-adjoint $p \equiv P_{D,N} - P_{D,N}$. Using $U(a)$, we can check the following inclusion property $N' = U(1)NU(-1)$, where $U(a) = e^{iap}$.

Here in the spacetime representation, 
\begin{align}
p_{(N,N')} = \frac{z_{31}(z_{2}-\zeta)^{2}}{z_{12}z_{32}}\partial_{\zeta} + \frac{z_{31}(z_{2}-\bar{\zeta})^{2}}{z_{12}z_{32}}\partial_{\bar{\zeta}}
\end{align}
Hence the action of $U(1)$ on spacetime point $(\zeta,\bar{\zeta})$ gives
\begin{align}
e^{p_{(N,N')}}(\zeta,\bar{\zeta}) = \left(\frac{\alpha z_{2}(\zeta-z_{2})+\zeta}{\alpha (\zeta-z_{2})+1},\frac{\alpha z_{2}(\bar{\zeta}-z_{2})+\bar{\zeta}}{\alpha (\bar{\zeta}-z_{2})+1}\right) \; ; \; \alpha = \frac{z_{31}}{z_{12}z_{32}}
\end{align}
Here this particular $SL(2,\mathbb{R})$ transformation $\zeta \rightarrow \frac{(\alpha z_{2}+1)\zeta- \alpha z_{2}^{2}}{\alpha \zeta +1-\alpha z_{2}}$ gives the map from the larger diamond $N$ to smaller diamond $N$. For instance, the left tip $(z_{3},z_{3})$ maps to $(z_{1},z_{1})$, upper tip $(z_{2},z_{3})$ maps to that of $N'$ i.e $(z_{2},z_{1})$ and so on. Using the reverse transformation $U(-1)$ one could construct $N$ from $N'$. Similarly we could treat $N'' \subset N$ as a -half sided $P_{D}$ inclusion as the commutators gives an overall minus sign. In the same way, one can define
\begin{align}
p_{(N,N'')} = \frac{z_{12}(z_{3}-\zeta)^{2}}{z_{32}z_{31}}\partial_{\zeta} + \frac{z_{12}(z_{3}-\bar{\zeta})^{2}}{z_{32}z_{31}}\partial_{\bar{\zeta}}
\end{align}
Here the action of $U(-1)$ is given by
\begin{align}
e^{-p_{(N,N'')}}(\zeta,\bar{\zeta}) = \left(\frac{(\beta z_{3} - 1)\zeta-\beta z_{3}^{2}}{\beta \zeta -\beta z_{3} - 1},\frac{(\beta z_{3} - 1)\bar{\zeta}-\beta z_{3}^{2}}{\beta \bar{\zeta} -\beta z_{3} - 1}\right) \; ; \; \beta = \frac{z_{12}}{z_{32}z_{31}}
\end{align}
In this map $z_{2}\rightarrow z_{1}$ and $z_{3}$ remains unchanged and thus it transforms $N$ to $N''$. Hence using $p_{N,N'}$, $p_{N,N''}$ consecutively we can map $N'$ to $N''$ and vice versa.  In this way, $P_{D}$ inclusion gives a natural way to map between diamonds with the structures like $N$, $N'$, $N''$. 
\vskip 0.2 cm
\textbf{$K$-inclusion}\\
Let us look at the another set of algebra described in (\ref{k inclusion}) which provides another notion of modular inclusion which we call `$K$-modular inclusion'. It consists of three diamonds $N$, $U$, $L$ such that $U$, $L \subset N$. Since the total modular Hamiltonian $K$ generates a flow from lower to upper tip, the algebra of $U$ and $L$ left unchanged under the flow of $K^{N}$, i.e $e^{-iK^{N}t}(U,L)e^{iK^{N}t} \subset (U,L)$. In a similar way of $P_{D}$-inclusion, here we can define a self adjoint $p \equiv K^{N}-K^{U,L}$. For instance, considering the inclusion $U\subset N$, we have
\begin{align}
\frac{z_{12}(z_{3}-\zeta)^{2}}{z_{32}z_{31}}\partial_{\zeta} + \frac{z_{13}(z_{2}-\bar{\zeta})^{2}}{z_{12}z_{32}}\partial_{\bar{\zeta}}
\end{align} 
Hence the action of $U(1)$ gives,
\begin{align}
e^{p_{(N,U)}}(\zeta,\bar{\zeta}) = \left(\frac{(\beta z_{3} - 1)\zeta-\beta z_{3}^{2}}{\beta \zeta- \beta z_{3} - 1},\frac{\alpha z_{2}(\bar{\zeta}-z_{2})+\bar{\zeta}}{\alpha (\bar{\zeta}-z_{2})+1}\right) 
\end{align}
In this map, one could see the left tip of $N$ i.e $(z_{3},z_{3})$ maps to left tip of $U$ i.e $(z_{3},z_{1})$, the right tip $(z_{2},z_{2})$ of $N$ maps to that of $U$ i.e $(z_{1},z_{2})$, the lower tip $(z_{2},z_{3})$ maps to the same $(z_{1},z_{1})$ and the upper tip remains unchanged for both diamonds. In the similar fashion, we could obtain the map from $N$ to $L$ using -half-sided $K$-inclusion of $L \subset N$. 

Also both $K$-inclusion and $P_{D}$-inclusion of the form (\ref{k inclusion}) and (\ref{p inclusion}), satisfy $SL(2,\mathbb{R})$ algebra which we describe above as a theorem.

Using the fact that any modular generators of any diamond can be constructed from the modular algebra of $N'$, $N$, $N''$ and using the above mentioned $K$ and $P_{D}$ inclusion, we can now reproduce all causal diamonds and the fields of them from the modular conformal generators.

\subsection{Modular inclusion in finite dimensional Hilbert space}
%\textcolor{red}{Somnath} and \textcolor{green}{Baishali} please rewrite this section briefly in a logical manner. First describe the system you took and argue what condition \ref{+hsmi} imposes on the system. Then describe briefly the set up and discuss what you get from \ref{inclusion equation}.
Let us consider a finite dimensional quantum system and divide it into four subsystems $A,A',B $and $B'$, such that the dimensions of subsystems are related in the following way:
\begin{align}
H_{tot}=H_{A}\otimes H_{A'}\otimes H_{B}\otimes H_{B'}; \; d_A=d_{A'}=N, d_B=d_{B'}=N'.
\end{align}
Without any loss of generality, we also assume that the total Hilbert space can be factorized as $H_{tot}=H_{AA'}\otimes H_{BB'}$, such that, there exists the state vectors $\ket{\psi}\in H_{tol}$, $\ket{\phi}\in H_{AA'}$ and $\ket{\chi}\in H_{BB'}$ which satisfy 
\begin{equation*}
\ket{\psi}=\ket{\phi}_{AA'}\otimes\ket{\chi}_{BB'}
\end{equation*}
We will first show that for such construction of the state $\ket{\psi}$, the modular inclusion criterion (\ref{+hsmi}) will be automatically satisfied. Here we take $M$ to be the system $AB$ and $\tilde{M}$ to be $A$. To show this, we first define the corresponding density matrices and reduced density matrices as follows:
\begin{align}
&\rho_{\psi} = \ket{\phi}\bra{\phi}\otimes \ket{\chi}\bra{\chi} \\
&\rho_{\!_{AB}}=tr_{\!_{A'B'}}\rho =tr_{A'}\ket{\phi}\bra{\phi}\otimes tr_{B'} \ket{\chi}\bra{\chi} \\
&\rho_{\!_{A'B'}}=tr_{\!_{AB}}\rho =tr_{A}\ket{\phi}\bra{\phi}\otimes tr_{B} \ket{\chi}\bra{\chi} \\
&\rho_{\!_{A}}=tr_{\!_{A'BB'}}\rho =tr_{A'}\ket{\phi}\bra{\phi} \\
&\rho_{\!_{A^\complement}}=tr_{\!_{A}}\rho =tr_{A}\ket{\phi}\bra{\phi}\otimes  \ket{\chi}\bra{\chi}
\end{align}
The total modular Hamiltonians $K_{M,\tilde{M}}$ or the modular operator $\Delta_{M,\tilde{M}}$ for the regions $M$ and $\tilde{M}$ are defined as:
\begin{equation}\label{Deltaab}
    \Delta_{M}\equiv \Delta_{AB}=\rho_{\!_{AB}}\otimes{\rho_{\!_{A'B'}}}^{-1}
\end{equation}
\begin{equation}\label{Deltaa}
    \Delta_{N}\equiv\Delta_{A}=\rho_{\!_{A}}\otimes{\rho_{\!_{A^\complement}}}^{-1}
\end{equation}
\vskip 0.09 cm
To begin with, we use Schmidt decomposition of $\ket{\phi}_{AA'}$ and $\ket{\chi}_{BB'}$ as following:
\begin{equation}
    \ket{\phi}_{AA'}=\sum \limits_{i=1}^N C_{i} \ket{i}_{A}\otimes\ket{i}_{A'}
\end{equation}
And
\begin{equation}
    \ket{\chi}_{BB'}=\sum \limits_{k=1}^{N'} D_{k} \ket{k}_{B}\otimes\ket{k}_{B'}
\end{equation}
Hence in this basis, we get
\begin{equation*}
    \rho_{\!_{AB}}=\sum \limits_{i,k=1}^{N,N'} \abs {C_{i}}^2\abs {D_{k}}^2\ket{i}_{A}\ket{k}_{B}\bra{i}_{A}\bra{k}_{B}
\end{equation*}
\begin{equation*}
    \rho_{\!_{A'B'}}=\sum \limits_{i,k} \abs {C_{i}}^2\abs {D_{k}}^2\ket{i}_{A'}\ket{k}_{B'}\bra{i}_{A'}\bra{k}_{B'}
\end{equation*}
Using the definition of (\ref{Deltaab}), we get:
\begin{equation}
 {\Delta_{AB}}^{it} =  (\, \sum \limits_{i,j,k,l} \frac{{\abs {C_{i}}^2\abs {D_{j}}^2}}{\abs {C_{k}}^2\abs {D_{l}}^2})\,^{it}(\ket{i}_{A}\ket{j}_{B}\bra{i}_{A}\bra{j}_{B})(\ket{k}_{A'}\ket{l}_{B'}\bra{k}_{A'}\bra{l}_{B'}).
\end{equation}
To show the inclusion condition (\ref{+hsmi}), we need to define an operator which has support only in the region A, as :
\begin{equation}
    \sum_{m,n} \mathcal{O}_{m,n}\ket{m}_{A}\bra{n}_{A}\otimes\mathbb{I}_{\!_{A'}}\otimes\mathbb{I}_{\!_{B}}\otimes\mathbb{I}_{\!_{B'}}
\end{equation}
Using the definition of $\Delta$, it is straightforward to show that,
\begin{equation}
    {\Delta_{AB}}^{-it}(\sum_{m,n} \mathcal{O}_{m,n}\ket{m}_{A}\bra{n}_{A}\otimes\mathbb{I}_{\!_{A'}}\otimes\mathbb{I}_{\!_{B}}\otimes\mathbb{I}_{\!_{B'}}){\Delta_{AB}}^{it}=\sum_{m,n}(\frac{{\abs {C_{m}}^2}}{\abs {C_{n}}^2})^{it} \mathcal{O}_{m,n}\ket{m}_{A}\bra{n}_{A}\otimes\mathbb{I}_{\!_{A'}}
    \otimes\mathbb{I}_{\!_{B}}\otimes\mathbb{I}_{\!_{B'}}
\end{equation}
From the above equation, it is clear that the state $\psi$ satisfy the equation (\ref{+hsmi}). With this, we want to check explicitly if it satisfies the condition of (\ref{inclusion equation}). To do so, we need to evaluate $\rho_{\!_{A^\complement}}$. However, $\rho_{\!_{A^{\!_{\complement}}}}^{-1}$ may not be defined. Since we are calculating  $ln\Delta_{A}$, this won't matter.
We can write,
\begin{align}
   & \ln\Delta_{AB}=\ln\rho_{\!_{AB}}\otimes{ \mathbb{I}_{\!_{A'B'}} }-\mathbb{I}_{\!_{AB}}\otimes{\ln\rho_{\!_{A'B'}}}\\
   & \ln\Delta_{A}=\ln\rho_{\!_{A}}\otimes{ \mathbb{I}_{\!_{A^{\complement}}} }-\mathbb{I}_{\!_{A}}\otimes{\ln\rho_{\!_{A^{\complement}}}}
\end{align}
To calculate the commutation, we act $\ln\Delta_{AB}$ and $\ln\Delta_{A}$ consecutively on a basis state $\ket{i}_{A}\ket{j}_{A'}\ket{i}_{B}\ket{j}_{B'}$ . One can see that,
$\ln\Delta_{AB}\ket{i}_{A}\ket{j}_{A'}\ket{i}_{B}\ket{j}_{B'}=0$
So,
\begin{equation}
    \bra{j'}_{B'}\bra{i'}_{B}\bra{j'}_{A'}\bra{i'}_{A}(\ln\Delta_{A}\ln\Delta_{AB})\ket{i}_{A}\ket{j}_{A'}\ket{i}_{B}\ket{j}_{B'}=0
\end{equation}
In the similar manner we can also check that,
\begin{equation*}
    \ln\Delta_{A}\ket{i}_{A}\ket{j}_{A'}\ket{i}_{B}\ket{j}_{B'}= (\ln\abs{C_{i}}^2 - \ln\abs{C_{j}}^2)\ket{i}_{A}\ket{j}_{A'}\ket{i}_{B}\ket{j}_{B'} 
\end{equation*} 
Since $\ln\Delta_{AB}\ket{i}_{A}\ket{j}_{A'}\ket{i}_{B}\ket{j}_{B'}=0$, it follows from above that,
\begin{equation}
    \bra{j'}_{B'}\bra{i'}_{B}\bra{j'}_{A'}\bra{i'}_{A}\ln\Delta_{AB}\ln\Delta_{A}\ket{i}_{A}\ket{j}_{A'}\ket{i}_{B}\ket{j}_{B'}=0
\end{equation}
So from the above equations, we finally get
\begin{equation}
    \bra{j'}_{B'}\bra{i'}_{B}\bra{j'}_{A'}\bra{i'}_{A}[\ln\Delta_{AB},\ln\Delta_{A}]\ket{i}_{A}\ket{j}_{A'}\ket{i}_{B}\ket{j}_{B'}= 0
\end{equation}
Similarly, one can easily check that
\begin{equation}
   \bra{j'}_{B'}\bra{i'}_{B}\bra{j'}_{A'}\bra{i'}_{A}\left(\ln\Delta_{AB}-\ln\Delta_{A}\right)\ket{i}_{A}\ket{j}_{A'}\ket{i}_{B}\ket{j}_{B'}=0
\end{equation}
Therefore in such example we get the desired inclusion properties (since it is true for any basis state)
\begin{equation*}
[\ln\Delta_{AB},\ln\Delta_{A}]=\ln\Delta_{AB}-\ln\Delta_{A}
\end{equation*}
Thus for such finite dimensional quantum system modular inclusion still holds.
%\textcolor{red}{Somnath} and \textcolor{green}{Baishali} please rewrite this section briefly in a logical manner. First describe the system you took and argue what condition \ref{+hsmi} imposes on the system. Then describe briefly the set up and discuss what you get from \ref{inclusion equation}.
\section{Commutation relation of modular generators and Virasoro algebra}\label{B}
Here we will reproduce the Virasoro algebra (\ref{Virasoro}) from the expression of $L_{n}$s which is of the form (\ref{Ln definition}), using the commutation relations of stress energy tensors. In CFT$_{2}$, $TT$ OPE takes the following form,
\begin{align}
T(z)T(\omega) = \frac{c/2}{(z-\omega)^{4}}+ \frac{2T(\omega)}{(z-\omega)^{2}}+\frac{\partial T(\omega)}{z-\omega}+\text{regular terms}
\end{align}
Using the Sokhotski-Plemelj formula, after analytically continuing to lightcone coordinate by $i\epsilon$ prescription, we get the following stress tensor commutators which we need to evaluate the $L_{n}$ commutators.
\begin{align}
[T(\zeta),T(\omega)] = 2\pi i [ -\frac{c}{12}\partial^{3}_{\omega}\delta(\omega-\zeta)+\delta(\omega-\zeta)\partial_{\omega}T(\omega)+2\partial_{\omega}\delta(\omega-\zeta)T(\omega)]\\
[\bar{T}(\bar{\zeta}),\bar{T}(\bar{\omega})] = -2\pi i [ -\frac{c}{12}\partial^{3}_{\bar{\omega}}\delta(\bar{\omega}-\bar{\zeta})+\delta(\bar{\omega}-\bar{\zeta})\partial_{\bar{\omega}}\bar{T}(\bar{\omega})+2\partial_{\bar{\omega}}\delta(\bar{\omega}-\bar{\zeta})\bar{T}(\bar{\omega})]
\end{align}
Inserting this into the commutator of $\mathbb{L}_{m}$, we have
\begin{align}
&[\mathbb{L}_{m} , \mathbb{L}_{n}] \nonumber \\
&=\int^{\infty}_{-\infty} d\zeta \int^{\infty}_{-\infty}d\omega \frac{(z_{2}-\zeta)^{-m+1}(\zeta-z_{3})^{m+1}}{(z_{2}-z_{3})^{2}}\left(\frac{z_{21}}{z_{31}}\right)^{m+n}(z_{2}-\omega)^{-n+1}(\omega-z_{3})^{n+1} \nonumber \\
& \times 2\pi i [- \frac{c}{12}\partial^{3}_{\omega}\delta(\omega-\zeta)+\delta(\omega-\zeta)\partial_{\omega}T(\omega)+2\partial_{\omega}\delta(\omega-\zeta)T(\omega)]
\end{align}
First let us consider the last two term (ignoring the $c$ term) of the $[T,T]$, we have
\begin{align}
& \frac{1}{2\pi i}[\mathbb{L}_{m} , \mathbb{L}_{n}]^{(1)} \nonumber \\
& = - \int^{\infty}_{-\infty} d\zeta \frac{(z_{2}-\zeta)^{-m-n+2}(\zeta-z_{3})^{m+n+2}}{(z_{2}-z_{3})^{2}}\left(\frac{z_{21}}{z_{31}}\right)^{m+n}\partial_{\zeta}T(\zeta) + 2\text{(T.D)}_{1}\nonumber \\
& -2\int^{\infty}_{-\infty} d\zeta \frac{(z_{2}-\zeta)^{-m+1}(\zeta-z_{3})^{m+1}}{(z_{2}-z_{3})^{2}}\left(\frac{z_{21}}{z_{31}}\right)^{m+n}T(\zeta)\nonumber \\ 
&\times  \left[(n-1)(z_{2}-\zeta)^{-n}(\zeta-z_{3})^{n+1}+ (n+1)(z_{2}-\zeta)^{-n+1}(\zeta-z_{3})^{n}\right] \nonumber \\
&= 2\text{(T.D)}_{1} - \text{(T.D)}_{2} \nonumber \\
&+\int^{\infty}_{-\infty}d\zeta \frac{(z_{2}-\zeta)^{-m-n+1}(\zeta-z_{3})^{m+n+1}}{(z_{2}-z_{3})^{2}}\left(\frac{z_{21}}{z_{31}}\right)^{m+n} \left[(m+n-2)(\zeta-z_{3})+(m+n+2)(z_{2}-\zeta)\right]T(\zeta)\nonumber \\
& - \int^{\infty}_{-\infty}d\zeta \frac{(z_{2}-\zeta)^{-m-n+1}(\zeta-z_{3})^{m+n+1}}{(z_{2}-z_{3})^{2}}\left(\frac{z_{21}}{z_{31}}\right)^{m+n} \left[(2n-2)(\zeta-z_{3})+(2n+2)(z_{2}-\zeta)\right]T(\zeta)\nonumber \\
&=2\text{(T.D)}_{1} - \text{(T.D)}_{2} + (m-n) \int^{\infty}_{-\infty}d\zeta  \frac{(z_{2}-\zeta)^{-m-n+1}(\zeta-z_{3})^{m+n+1}}{z_{2}-z_{3}}\left(\frac{z_{21}}{z_{31}}\right)^{m+n}T(\zeta)\zeta
\end{align}
We can identify the last term as the $(m-n)\mathbb{L}_{m+n}$. Here $(T.D)_{1,2}$ are two total derivative terms coming from the intermediate steps of the partial integration. Here,
\begin{align}
&\text{(T.D)}_{1} =  \int^{\infty}_{-\infty}d\zeta  \frac{(z_{2}-\zeta)^{-m+1}(\zeta-z_{3})^{m+1}}{(z_{2}-z_{3})^{2}}\left(\frac{z_{21}}{z_{31}}\right)^{m+n}  \bigg[(z_{2}-\omega)^{-n+1}(\omega-z_{3})^{n+1}\delta(\omega-\zeta)T(\omega)\bigg]^{\omega=\infty}_{\omega=-\infty}
\end{align} 
Due to the Dirac delta function, the total derivative term inside the bracket, vanishes. Hence this term vanishes. The other term is,
\begin{align}
&\text{(T.D)}_{2} = \left(\frac{z_{21}}{z_{31}}\right)^{m+n}\left[  \frac{(z_{2}-\zeta)^{-m-n+2}(\zeta-z_{3})^{m+n+2}}{(z_{2}-z_{3})^{2}}T(\zeta)\right]^{\zeta = \infty}_{\zeta = -\infty}
\end{align}
To analyze this term, we need to look at the behavior of the stress tensor near spacetime infinity. From the transformation property of stress tensor we know, $T'(\zeta') = \left(\frac{\partial z'}{\partial z}\right)^{-2}+$ Schwarzian derivative term. We choose a global transformation\footnote{Hence we can get rid of the Schwarzian derivative term.} $\zeta' = \frac{a\zeta+b}{c\zeta+d}$ at $\zeta = \zeta_{0} = -\frac{d}{c}+\epsilon$, such that $\zeta'_{0} \sim \frac{1}{\epsilon}$. For such choice, we get the transformation of stress tensor in the following way,
\begin{align}
T'(\zeta'_{0}) = (c\zeta_{0}+d)^{4}T(\zeta_{0}) \sim \frac{1}{\epsilon^{4}}T(\zeta_{0})
\end{align} 
Hence, for $\epsilon \rightarrow 0$, we get the behavior of the stress tensor near infinity as $T(\zeta)|_{\zeta\rightarrow\infty} \sim \frac{1}{\zeta^{4}}$. Using this, if we look at the term $(T.D)_{2}$, we get,
\begin{align}
&\text{(T.D)}_{2} = \left(\frac{z_{21}}{z_{31}}\right)^{m+n}\lim_{\Lambda\rightarrow\infty}\left[  \frac{(z_{2}-\Lambda)^{-m-n+2}(\Lambda-z_{3})^{m+n+2}}{(z_{2}-z_{3})^{2}}\frac{1}{\Lambda^{4}}-\Big(\Lambda \rightarrow - \Lambda\Big)\right]=0
\end{align}
Hence, both the total derivative terms vanishes. Let us now see the contribution coming from the central charge($c$) part of the stress tensor commutator. 
\begin{align}
& \frac{1}{2\pi i}[\mathbb{L}_{m} , \mathbb{L}_{n}]^{(2)} \nonumber \\
& = -\frac{c}{12}\left(\frac{z_{21}}{z_{31}}\right)^{m+n}\int^{\infty}_{-\infty} d\zeta \int^{\infty}_{-\infty}d\omega \frac{(z_{2}-\zeta)^{-m+1}(\zeta-z_{3})^{m+1}}{(z_{2}-z_{3})^{2}}(z_{2}-\omega)^{-n+1}(\omega-z_{3})^{n+1} \partial^{3}_{\omega}\delta(\omega-\zeta)
\end{align}
Let us denote the constant term $\frac{c}{12}\left(\frac{z_{21}}{z_{31}}\right)^{m+n}\frac{1}{(z_{2}-z_{3})^{2}} \equiv A$. In a similar fashion of the previous calculation, after some simple algebraic steps the final integration is of the following form
\begin{align}
& \frac{1}{2\pi i}[\mathbb{L}_{m} , \mathbb{L}_{n}]^{(2)} \nonumber \\
& = \text{(T.D)}_{3} + \text{(T.D)}_{4} + \text{(T.D)}_{5} + n(n^{2}-1)A(z_{2}-z_{3})^{3} \int^{\infty}_{-\infty}d\zeta (z_{2}-\zeta)^{-m-n-1}(\zeta-z_{3})^{m+n-1}
\end{align}
Here, the total derivative terms $(\text{T.D})_{3,4,5}$ are getting vanished due to the presence of dirac delta function and it's derivatives as we argued before. After carefully choosing a contour, we get the final result of the complex integration as (we choose $\text{Re}[z_{2}]>0, \text{Re}[z_{3}]<0 $)
\begin{align}
& \frac{1}{2\pi i}[\mathbb{L}_{m} , \mathbb{L}_{n}]^{(2)} = \frac{c}{12}n(n^{2}-1)\left(\frac{z_{21}}{z_{31}}\right)^{m+n} \frac{\left(\frac{-z_{2}}{z_{2}}\right)^{m+n}-\left(\frac{-z_{3}}{z_{3}}\right)^{m+n}}{m+n};  
\end{align}
This term vanishes for any $m+n \neq 0, \in \mathbb{Z}$. To extract the contribution for $m+n = 0$, we can perform an analytic continuation by choosing $m+n = \epsilon$, taking $\epsilon \rightarrow 0$. This gives,
\begin{align}
 \frac{1}{2\pi i}[\mathbb{L}_{m} , \mathbb{L}_{n}]^{(2)} = \frac{c}{12}n(n^{2}-1)\lim_{\epsilon \rightarrow 0}\left(\frac{z_{21}}{z_{31}}\right)^{\epsilon} \frac{(-1)^{\epsilon}-(-1)^{-\epsilon}}{\epsilon} = \frac{c}{12}n(n^{2}-1)2\pi i 
\end{align}
Hence, combining $[\mathbb{L}_{m} , \mathbb{L}_{n}]^{(1)}$ and $[\mathbb{L}_{m} , \mathbb{L}_{n}]^{(2)}$, we finally have
\begin{align}
[\mathbb{L}_{m},\mathbb{L}_{n}] = (m-n)\mathbb{L}_{m+n} + \frac{c}{12}n(n^{2}-1)\delta_{m+n,0}
\end{align}
Here, $\mathbb{L}_{m,n}$s are redefined as $\mathbb{L}_{m,n} \rightarrow \frac{1}{2\pi i}\mathbb{L}_{m,n}$.
%\section{Modular inclusion in finite dimension}\label{C}


\begin{thebibliography}{99}

\bibitem{Haag:1992hx}
R.~Haag,
``Local quantum physics: Fields, particles, algebras,''

\bibitem{Witten:2018zxz}
E.~Witten,
``APS Medal for Exceptional Achievement in Research: Invited article on entanglement properties of quantum field theory,''
Rev. Mod. Phys. \textbf{90}, no.4, 045003 (2018)

\bibitem{Casini:2008cr}
H.~Casini,
``Relative entropy and the Bekenstein bound,''
Class. Quant. Grav. \textbf{25}, 205021 (2008)

\bibitem{Bousso:2014sda}
R.~Bousso, H.~Casini, Z.~Fisher and J.~Maldacena,
``Proof of a Quantum Bousso Bound,''
Phys. Rev. D \textbf{90}, no.4, 044002 (2014)

\bibitem{Bousso:2014uxa}
R.~Bousso, H.~Casini, Z.~Fisher and J.~Maldacena,
``Entropy on a null surface for interacting quantum field theories and the Bousso bound,''
Phys. Rev. D \textbf{91}, no.8, 084030 (2015)

\bibitem{Casini:2004bw}
H.~Casini and M.~Huerta,
``A Finite entanglement entropy and the c-theorem,''
Phys. Lett. B \textbf{600}, 142-150 (2004)

\bibitem{Casini:2012ei}
H.~Casini and M.~Huerta,
``On the RG running of the entanglement entropy of a circle,''
Phys. Rev. D \textbf{85}, 125016 (2012)

\bibitem{Casini:2016udt}
H.~Casini, E.~Teste and G.~Torroba,
``Relative entropy and the RG flow,''
JHEP \textbf{03}, 089 (2017)

\bibitem{Casini:2017vbe}
H.~Casini, E.~Test\'e and G.~Torroba,
``Markov Property of the Conformal Field Theory Vacuum and the a Theorem,''
Phys. Rev. Lett. \textbf{118}, no.26, 261602 (2017)

\bibitem{Casini:2018nym}
H.~Casini, I.~Salazar Landea and G.~Torroba,
``Irreversibility in quantum field theories with boundaries,''
JHEP \textbf{04}, 166 (2019)

\bibitem{Faulkner:2016mzt}
T.~Faulkner, R.~G.~Leigh, O.~Parrikar and H.~Wang,
``Modular Hamiltonians for Deformed Half-Spaces and the Averaged Null Energy Condition,''
JHEP \textbf{09}, 038 (2016)

\bibitem{Koeller:2017njr}
J.~Koeller, S.~Leichenauer, A.~Levine and A.~Shahbazi-Moghaddam,
``Local Modular Hamiltonians from the Quantum Null Energy Condition,''
Phys. Rev. D \textbf{97}, no.6, 065011 (2018)

\bibitem{Balakrishnan:2017bjg}
S.~Balakrishnan, T.~Faulkner, Z.~U.~Khandker and H.~Wang,
``A General Proof of the Quantum Null Energy Condition,''
JHEP \textbf{09}, 020 (2019)

\bibitem{Ceyhan:2018zfg}
F.~Ceyhan and T.~Faulkner,
``Recovering the QNEC from the ANEC,''
Commun. Math. Phys. \textbf{377}, no.2, 999-1045 (2020)


\bibitem{Jafferis:2015del}
D.~L.~Jafferis, A.~Lewkowycz, J.~Maldacena and S.~J.~Suh,
``Relative entropy equals bulk relative entropy,''
JHEP \textbf{06}, 004 (2016)

\bibitem{Faulkner:2017vdd}
T.~Faulkner and A.~Lewkowycz,
``Bulk locality from modular flow,''
JHEP \textbf{07}, 151 (2017)

\bibitem{Kabat:2017mun}
D.~Kabat and G.~Lifschytz,
``Local bulk physics from intersecting modular Hamiltonians,''
JHEP \textbf{06}, 120 (2017)

\bibitem{Sarosi:2017rsq}
G.~S\'arosi and T.~Ugajin,
``Modular Hamiltonians of excited states, OPE blocks and emergent bulk fields,''
JHEP \textbf{01}, 012 (2018)

\bibitem{Das:2018ojl}
S.~Das and B.~Ezhuthachan,
``Modular Hamiltonians and large diffeomorphisms in AdS$_{3}$,''
JHEP \textbf{12}, 096 (2018)

\bibitem{Czech:2017zfq}
B.~Czech, L.~Lamprou, S.~Mccandlish and J.~Sully,
``Modular Berry Connection for Entangled Subregions in AdS/CFT,''
Phys. Rev. Lett. \textbf{120}, no.9, 091601 (2018)

\bibitem{Faulkner:2018faa}
T.~Faulkner, M.~Li and H.~Wang,
``A modular toolkit for bulk reconstruction,''
JHEP \textbf{04}, 119 (2019)

\bibitem{Czech:2019vih}
B.~Czech, J.~De Boer, D.~Ge and L.~Lamprou,
``A modular sewing kit for entanglement wedges,''
JHEP \textbf{11}, 094 (2019)

\bibitem{deBoer:2019uem}
J.~De Boer and L.~Lamprou,
``Holographic Order from Modular Chaos,''
JHEP \textbf{06}, 024 (2020)

\bibitem{Blanco:2013joa}
D.~D.~Blanco, H.~Casini, L.~Y.~Hung and R.~C.~Myers,
``Relative Entropy and Holography,''
JHEP \textbf{08}, 060 (2013)

\bibitem{Lashkari:2016idm}
N.~Lashkari, J.~Lin, H.~Ooguri, B.~Stoica and M.~Van Raamsdonk,
``Gravitational positive energy theorems from information inequalities,''
PTEP \textbf{2016}, no.12, 12C109 (2016)

\bibitem{Blanco:2017akw}
D.~Blanco, H.~Casini, M.~Leston and F.~Rosso,
``Modular energy inequalities from relative entropy,''
JHEP \textbf{01}, 154 (2018)

\bibitem{Faulkner:2013ica}
T.~Faulkner, M.~Guica, T.~Hartman, R.~C.~Myers and M.~Van Raamsdonk,
``Gravitation from Entanglement in Holographic CFTs,''
JHEP \textbf{03}, 051 (2014)

\bibitem{Faulkner:2017tkh}
T.~Faulkner, F.~M.~Haehl, E.~Hijano, O.~Parrikar, C.~Rabideau and M.~Van Raamsdonk,
``Nonlinear Gravity from Entanglement in Conformal Field Theories,''
JHEP \textbf{08}, 057 (2017)

\bibitem{Roy:2018ehv}
S.~R.~Roy and D.~Sarkar,
``Bulk metric reconstruction from boundary entanglement,''
Phys. Rev. D \textbf{98}, no.6, 066017 (2018)


\bibitem{Kabat:2018smf}
D.~Kabat and G.~Lifschytz,
``Emergence of spacetime from the algebra of total modular Hamiltonians,''
JHEP \textbf{05}, 017 (2019)

\bibitem{Casini:2011kv}
H.~Casini, M.~Huerta and R.~C.~Myers,
``Towards a derivation of holographic entanglement entropy,''
JHEP \textbf{05}, 036 (2011)

\bibitem{Bousso:2020yxi}
R.~Bousso, V.~Chandrasekaran, P.~Rath and A.~Shahbazi-Moghaddam,
``Gravity dual of Connes cocycle flow,''
Phys. Rev. D \textbf{102}, no.6, 066008 (2020)

\bibitem{Levine:2020upy}
A.~Levine, A.~Shahbazi-Moghaddam and R.~M.~Soni,
``Seeing the entanglement wedge,''
JHEP \textbf{06}, 134 (2021)

\bibitem{Czech:2016xec}
B.~Czech, L.~Lamprou, S.~McCandlish, B.~Mosk and J.~Sully,
``A Stereoscopic Look into the Bulk,''
JHEP \textbf{07}, 129 (2016)

\bibitem{deBoer:2016pqk}
J.~de Boer, F.~M.~Haehl, M.~P.~Heller and R.~C.~Myers,
``Entanglement, holography and causal diamonds,''
JHEP \textbf{08}, 162 (2016)

\bibitem{Czech:2015qta}
B.~Czech, L.~Lamprou, S.~McCandlish and J.~Sully,
``Integral Geometry and Holography,''
JHEP \textbf{10}, 175 (2015)

\bibitem{Das:2019iit}
S.~Das and B.~Ezhuthachan,
``Spectrum of Modular Hamiltonian in the Vacuum and Excited States,''
JHEP \textbf{10}, 009 (2019)

\bibitem{Das:2018ajg}
S.~Das,
``Comments on spinning OPE blocks in AdS$_{3}$/CFT$_{2}$,''
Phys. Lett. B \textbf{792}, 397-405 (2019)

\bibitem{Ferrara:1973vz} 
  S.~Ferrara, A.~F.~Grillo, G.~Parisi and R.~Gatto,
  ``Covariant expansion of the conformal four-point function,''
  Nucl.\ Phys.\ B {\bf 49}, 77 (1972)
  Erratum: [Nucl.\ Phys.\ B {\bf 53}, 643 (1973)].
  
 \bibitem{SimmonsDuffin:2012uy} 
  D.~Simmons-Duffin,
  ``Projectors, Shadows, and Conformal Blocks,''
  JHEP {\bf 1404}, 146 (2014)

\bibitem{Banados:1998gg}
M.~Banados,
``Three-dimensional quantum geometry and black holes,''
AIP Conf. Proc. \textbf{484}, no.1, 147-169 (1999)

\bibitem{Anand:2017dav}
N.~Anand, H.~Chen, A.~L.~Fitzpatrick, J.~Kaplan and D.~Li,
``An Exact Operator That Knows Its Location,''
JHEP \textbf{02}, 012 (2018)

\bibitem{Casini:2009vk}
H.~Casini and M.~Huerta,
``Reduced density matrix and internal dynamics for multicomponent regions,''
Class. Quant. Grav. \textbf{26}, 185005 (2009)

\bibitem{Erdmenger:2020nop}
J.~Erdmenger, P.~Fries, I.~A.~Reyes and C.~P.~Simon,
``Resolving modular flow: a toolkit for free fermions,''
JHEP \textbf{12}, 126 (2020)

\bibitem{Cardy:2016fqc}
J.~Cardy and E.~Tonni,
``Entanglement hamiltonians in two-dimensional conformal field theory,''
J. Stat. Mech. \textbf{1612}, no.12, 123103 (2016)

\bibitem{Apolo:2020qjm}
L.~Apolo, H.~Jiang, W.~Song and Y.~Zhong,
``Modular Hamiltonians in flat holography and (W)AdS/WCFT,''
JHEP \textbf{09}, 033 (2020)

\bibitem{Casini:2017roe}
H.~Casini, E.~Teste and G.~Torroba,
``Modular Hamiltonians on the null plane and the Markov property of the vacuum state,''
J. Phys. A \textbf{50}, no.36, 364001 (2017)

\bibitem{Jefferson:2018ksk}
R.~Jefferson,
``Comments on black hole interiors and modular inclusions,''
SciPost Phys. \textbf{6}, no.4, 042 (2019)


\bibitem{Borchers:1991xk}
H.~J.~Borchers,
``The CPT theorem in two-dimensional theories of local observables,''
Commun. Math. Phys. \textbf{143}, 315-332 (1992)

\bibitem{Wiesbrock:1992mg}
H.~W.~Wiesbrock,
``Half sided modular inclusions of von Neumann algebras,''
Commun. Math. Phys. \textbf{157}, 83-92 (1993)

\bibitem{Borchers:2000pv}
H.~J.~Borchers,
``On revolutionizing quantum field theory with Tomita's modular theory,''
J. Math. Phys. \textbf{41}, 3604-3673 (2000)

\bibitem{Wiesbrock:1996rq}
H.~W.~Wiesbrock,
``Symmetries and modular intersections of von Neumann algebras,''
Lett. Math. Phys. \textbf{39}, 203-212 (1997)

\bibitem{Wiesbrock:1996rp}
H.~W.~Wiesbrock,
``Modular intersections of von Neumann algebras in quantum field theory,''
Commun. Math. Phys. \textbf{193}, 269-285 (1998)
doi:10.1007/s002200050329

\bibitem{Kravchuk:2018htv}
P.~Kravchuk and D.~Simmons-Duffin,
``Light-ray operators in conformal field theory,''
JHEP \textbf{11}, 102 (2018)

\bibitem{Huang:2020ycs}
K.~W.~Huang,
``Lightcone Commutator and Stress-Tensor Exchange in $d>2$ CFTs,''
Phys. Rev. D \textbf{102}, no.2, 021701 (2020)

\bibitem{Belin:2020lsr}
A.~Belin, D.~M.~Hofman, G.~Mathys and M.~T.~Walters,
``On the Stress Tensor Light-ray Operator Algebra,''

\bibitem{Huang:2021hye}
K.~W.~Huang,
``$d>2$ Stress-Tensor OPE near a Line,''

\bibitem{Besken:2020snx}
M.~Besken, J.~de Boer and G.~Mathys,
``On Local and Integrated Stress-Tensor Commutators,''
[arXiv:2012.15724 [hep-th]].

\bibitem{Ishibashi:2015jba}
N.~Ishibashi and T.~Tada,
``Infinite circumference limit of conformal field theory,''
J. Phys. A \textbf{48}, no.31, 315402 (2015)

\bibitem{Ishibashi:2016bey}
N.~Ishibashi and T.~Tada,
``Dipolar quantization and the infinite circumference limit of two-dimensional conformal field theories,''
Int. J. Mod. Phys. A \textbf{31}, no.32, 1650170 (2016)
\end{thebibliography}
\end{document}